\documentclass{IEEEtran}
\usepackage[hidelinks]{hyperref}

\usepackage[printonlyused]{acronym}
\usepackage{amsmath}
\usepackage{balance}
\usepackage{url}
\usepackage{doi}
\usepackage{color}
\usepackage{graphicx}
\usepackage{cite}
\usepackage{array}

\usepackage[applemac]{inputenc}

\usepackage{ulem}
\usepackage{dblfloatfix}
\usepackage[ruled,norelsize]{algorithm2e}
\newcommand{\algoline}{}

\usepackage{mathtools}
\newcommand{\state}{x}
\newcommand{\mean}{\mu}

\newcommand{\augstate}{z}

\newcommand{\iterations}{j}
\newcommand{\statedim}{n}

\newcommand{\measfun}{h}
\newcommand{\meas}{y}

\newcommand{\part}[1]{\underline{#1}}

\newcommand{\statetransfun}{f}

\newcommand{\statenoise}{\varepsilon^\state}
\newcommand{\measnoise}[1][]{\varepsilon^\meas}

\newcommand{\statecov}{P}

\newcommand{\statefun}{f}

\newcommand{\sigmapw}{w}

\newcommand{\innocov}{S}
\newcommand{\kalmangain}{K}

\newcommand{\jacobian}{J}

\newcommand{\sigmap}{\chi}
\newcommand{\fun}{g}
\newcommand{\Mfun}{G}

    \providecommand{\matr}[1]{\begin{bmatrix} #1 \end{bmatrix}}
   \providecommand{\smatr}[1] {\begin{bmatrix} #1 \end{bmatrix} }

    \providecommand{\norm}[1]{\left|\hspace{-1pt}\left| {#1} \right|\hspace{-1pt}\right|}              
       
    \DeclareMathOperator{\E}{E}                                 % expectation value
    \DeclareMathOperator{\N}{N}                                 % normal distribution
	\newcommand{\noise}{\varepsilon}
    
    \newcommand{\nonlinearstatetrans}{f}
    \newcommand{\transnoise}{\noise_q}
    \newcommand{\Ex}[1]{\E\left[#1 \right]}
\newcommand{\tm}{t\!-\!1}

\newcommand{\Pd}{P^D_{xx,t_0}}
\newcommand{\Ph}{P_{\hat{x}\hat{x},t_1}}
\newcommand{\Pm}{P_{xx,t_0}}
\newcommand{\Rv}{ R_{t_1}^v}

\newcommand{\PdI}{ \left( { P^D_{xx,t_0} }\right)^{-1}}
\newcommand{\PhI}{ \left( {P_{\hat{x}\hat{x},t_1}}\right)^{-1}}
\newcommand{\PmI}{\left({P_{xx,t_0} }\right)^{-1}}
\newcommand{\RvI}{\left({R_{t_1}^v}\right)^{-1}}
 \DeclareMathOperator{\diag}{diag}

\begin{document}
\allowdisplaybreaks
    \newcounter{Uudet}
    \newcommand{\aref}[1]{\emph{Change}~\ref{#1} on page~\pageref{#1}}
    \newcommand{\uusi}[3][]{#3}

\newcommand{\U}[1]{\color{blue}#1\color{black}}

    \newcounter{Uudet2}
    \newcommand{\uus}[3][]{#3}
   \newcommand{\hp}[1]{}
\hp{
   
  \thispagestyle{empty}

We thank all the reviewers for their comments and suggestions of how to improve the manuscript. We have done changes based on the comments. Next there are the comments and our responses with pdf-links to changes. Main changes are shown with blue within the main body of the manuscript with a superscript number denoting the number of the change.

Responses to reviewer comments:

Reviewer \#1

\begin{itemize}
\item The English grammar needs to be improved. 

\emph{We fixed grammar based on other reviewers specific comments and read through the manuscript and fixed issues we found.}

\item  Since this is a tutorial more examples should be provided. Please provide an example that explicitly carries through the computations explicitly. Reduction in computational load should be contrasted with the solution as well. 

\emph{We added a new example in \aref{sec1} where we made more thorough evaluation of the complexity changes.}

\emph{We also added a second example in \aref{sec2} where complexity analysis shows that the optimizations are not worthwhile to implement in every situation.}

\item  Please add this reference for gains achieved by Strassen's matrix multiplication algorithm. 

\emph{We added the requested reference} \aref{strassen}
\end{itemize}

Reviewer \#2

\begin{itemize}
\item Many of the formulations focus on augmented state vectors that include parameters in the state vector that may not be of interest to estimate. The authors should at least mention that any easy way to improve efficiency it to not compute updates for state parameters that are not of interest. Certain parameters can be treated as nuisance parameters with their impact only account for in the state update covariance. Such is the case for the Schmidt Kalman Filter. 

\emph{We mentioned Schmidt-Kalman filter in \aref{schmidt} and provided a reference for an interested reader. The algorithm itself is not explained in detail as it is suboptimal.}

\item Introduction 2nd paragraph: "... applications the state may contain also variables ..."  switch the order of contain and also. 

\emph{We changed order in } \aref{alsoc}

\item 2. Introduction 5th paragraph:  ".... the estimate inaccurate when the true posterior ...."  Insert "is" between estimate and inaccurate 

\emph{We added is } \aref{addis}

\item  3. Last paragraph in section III  "blocks can be made in different rates"  replace in with at 

\emph{We replaced in with at in } \aref{inat}

\item 4. Equations 24 and 25 don't have the right brackets with the expected value and the brackets should be inside the E. e.g. E[ ] 

\emph{We changed the equations~\eqref{equ:U1}-\eqref{equ:U3} as requested}

\item 5. Section VI.A.  I don't like using y as the notation for the virtual update. y is initially presented as an observation. Why not just use x with the v superscript? 

\emph{As the virtual update is done as a Kalman update, $y^v$ corresponds to "virtual" observation. This is explicitly defined now in} \aref{virtual}

\item  6. Right before equation 76.  " ... pairs for the measurement noises are not independent ..." When e(i,j) = 0 don't you mean the measurements ARE independent because the correlation is 0? 

\emph{Our phrasing was incorrect. Actually the measurements would have been independent only if $e_i=0$ and $e_j=0$. If $e_{i,j}=0$ it means that the measurement error covariance is not full rank.  We rewrote the paragraph in} \aref{noisecov}.

\end{itemize}

Reviewer \#3
\begin{itemize}

\item "p1 KF is the optimal ..." include whiteness of noise. 

\emph{We included whiteness in} \aref{white} 
 
\item "p2 halving the dimension m ... reduces ..." by a factor of four would be slightly clearer. 

\emph{We changed the phrasing as requested in} \aref{byfour}

\item "p2 When state can be divided into decoupled blocks" incomplete sentence

\emph{We corrected the sentence in} \aref{inc}
 
\item "p2 The algorithms improve the computation efficiency" $\rightarrow$ "The algorithms improve in computation efficiency" 

\emph{We changed "the" $\rightarrow$ "in"} \aref{the}

\item "p6 If the blocks were independent at t0?" why the dagger? 

\emph{The dagger was a typo. We removed it}.

\item p8 Fix wrapping eqn numbers

\emph{We changed the equation formatting so that the equation numbers fit better on lines}
\end{itemize}
    \newpage
    \pagenumbering{arabic}
    \setcounter{page}{1}
}

\title{On computational complexity reduction methods for Kalman filter extensions}
\author{Matti Raitoharju${}^{\dagger\ddagger}$ and
 Robert Pich\'e${}^{\dagger}$\\
 ${}^\dagger$Faculty of Information Technology and Communication Sciences, Tampere University, Finland\\
 ${}^\ddagger$Department of Electrical Engineering and Automation, Aalto University, Finland} 
 
\maketitle

\begin{abstract}
The Kalman filter and its extensions are used in a vast number of aerospace and navigation applications for nonlinear state estimation of time series. In the literature, different approaches have been proposed to exploit the structure of the state and measurement models to reduce the computational demand of the algorithms. In this tutorial, we survey existing code optimization methods and present them using unified notation that allows them to be used with various Kalman filter extensions. We develop the optimization methods to cover a wider range of models, show how different structural optimizations can be combined, and present new applications for the existing optimizations. Furthermore, we present an example that shows that the exploitation of the structure of the problem can lead to improved estimation accuracy while reducing the computational load. This tutorial is intended for persons who are familiar with Kalman filtering and want to get insights for reducing the computational demand of different Kalman filter extensions. 

\end{abstract}
%\begin{keyword}{State estimation, Nonlinear systems, Computational complexity, Kalman filter}\end{keyword}

%\end{frontmatter}

\section{Introduction}

\begin{figure}[b!]
\footnotesize
This is an author accepted version of the  article. Final published article: \doi{10.1109/MAES.2019.2927898}.

© 2019 IEEE.  Personal use of this material is permitted.  Permission from IEEE must be obtained for all other uses, in any current or future media, including reprinting/republishing this material for advertising or promotional purposes, creating new collective works, for resale or redistribution to servers or lists, or reuse of any copyrighted component of this work in other works.
\end{figure}

Since its pioneering application to trajectory estimation in the Apollo program in the 1960's, the \ac{KF} and its nonlinear extensions have been used in a multitude of aerospace and navigation applications, including inertial navigation, radar systems, and global navigation satellite systems \cite{5466132}. \acp{KF} are also used in many other application areas, for example state estimation of a lithium polymer battery \cite{7525904} or brain imaging 
\cite{hiltunen2011state}. 
%
%In general, the Bayesian state estimate cannot be computed in closed form. Under certain conditions closed form solutions exist. For example, when measurement and state transition functions are linear and associated noises are Gaussian, the Bayesian state estimate can be computed in closed form by the \ac{KF} algorithm \cite{kalman}.  

\ac{KF} is the optimal Bayesian filter under certain conditions, which include linearity of models and Gaussianity \uus{white}{and whiteness} of noise \cite{kalman}.
 \acp{KFE} are based on approximate linear-Gaussian models that extend the use of \ac{KF} to nonlinear models. In the literature, there are several different types of \acp{KFE}, with different demands on computational resources. The computational complexity of the \ac{KFE} increases when the number of estimated state variables or dimensionality of the measurement vector increases; in some \acp{KFE} even exponentially \cite{ITO}. The number of state variables varies depending on the application, for example, in some positioning applications only 2 position variables are estimated, but in other positioning applications the state may \uus{alsoc}{also contain} variables for the locations of thousands of landmarks.  In many applications the computational resources are limited, for example, in miniaturized satellites \cite{6187240}. 

In this tutorial, we study various methods to reduce the computational load of state estimation with \acp{KFE} by exploiting the structure of the state transition and measurement models. This tutorial is intended for persons who are familiar with the basics of \acp{KFE} and want to know how to reduce the computational demand of \acp{KFE}. The presented algorithms are such that the result is exact when applied to a situation where the \ac{KF} produces the exact result. This leaves algorithms that are not optimal in the linear-Gaussian case, such as \ac{EnKF}, out of the scope of this tutorial. However, some of the given optimizations can still be applied with such algorithms.

We present the algorithms in a general form so that they can be applied to as wide a range of problems as possible, but still in a form that they are easy to implement. Some of the algorithms in the original sources are given only for a certain \ac{KFE}; the general notation used in this tutorial allows the optimizations to be implemented for different \acp{KFE}.  In addition to surveying the algorithms in the literature, we give some generalizations of the algorithms and new ways of applying the optimizations with the \acp{KFE}. To our knowledge,  there is no similar unified presentation of optimization methods in 
the literature.

A drawback of \acp{KFE} is that, because of the Gaussian approximation on which the algorithm is based, the estimate \uus{addis}{is} inaccurate when the true posterior is far from normal. \acp{GMF} (a.k.a. Gaussian sum filters)  use a sum of normal densities to estimate the probability distributions and can approximate any probability density function~\cite{Sorenson_Alspach_1971}. Because \acp{GMF} use several \acp{KFE} the computational load is larger than with algorithms that use only one Gaussian. The optimizations in this tutorial can be applied also in the \acp{GMF} for state propagation and update of individual components.

For implementations, we assume that a library for basic linear algebra is available. In many practical applications, the algorithms can be optimized further by taking the specific structure of the matrices into account. If the matrices are sparse, the sparsity can be exploited for optimization either by hand or by implementing the algorithms with sparsity-optimized subroutines. These can be found, for example, in Sparse Basic Linear Algebra Subprograms\footnote{\url{http://math.nist.gov/spblas/}} or in Matlab\footnote{\url{http://www.mathworks.com/products/matlab/}}. The optimizations in matrix algebra libraries also make it impossible to provide accurate complexity estimates of the given algorithms. For example the naive matrix multiplication for square matrices with $n$ columns has complexity $\mathcal{O}(n^3)$ while Strassen's algorithm has complexity $\mathcal{O}(n^{2.807})$ \uus{strassen}{\cite{strassen1969gaussian,golub2012matrix}} and algorithms with smaller complexity exist, although they are faster only with very large $n$.

The computation time of algorithms can also be reduced using parallel computing. The matrix operations can be effectively parallelized in the linear algebra library that handles the matrix operations. Thus, we do not consider parallelization in this tutorial.

The remainder of this tutorial is organized as follows. In the next section we present the common notations that are used throughout the tutorial. In Section~\ref{sec:background} the background of \acp{KFE} is presented. Section~\ref{sec:optimizations} contains  linearity based optimizations of the \acp{KFE}.  In Section~\ref{sec:inno} the solution of the set of linear equations containing the inverse of the innovation covariance is optimized. Section~\ref{sec:multiprocess} presents optimizations based on division of the state into multiple subprocesses. Section~\ref{sec:examples} gives example applications of the optimization methods. Section~\ref{sec:conclusions} concludes the article.

\section{Notations}
\label{sec:notations}
In the following list there are the variables that are used throughout the tutorial. In different sections, there are algorithm specific notations that are explained as they occur.\vspace{-0.5\topsep}
\begin{itemize}\renewcommand\labelitemi{--}
\item Scalars
\begin{description}
 	\item[$\iterations$] number of iterations
	\item[$m$] dimension of measurement
 	\item[$\statedim$] dimension of state (or augmented state)
 \end{description}
\item Random variables
\begin{description}
	\item[$\state$] state
	\item[$\augstate$] augmented state
	\item[$\noise$]  noise
	\item[$\statenoise$] state transition noise
	\item[$\measnoise$] measurement noise
\end{description}
\item Subscripts
\begin{description} 
 	\item[$i$] iteration index
 	\item[$t$] time index
\end{description}

\item Functions
\begin{description} 
	\item[$\statetransfun(\cdot)$] state transition function
	\item[$\fun(\cdot)$] non-specific function
	\item[$\Mfun(\cdot)$] Matrix valued function
	\item[$\measfun(\cdot)$] measurement function
	\item[$\diag(\cdot)$] diagonal matrix with function arguments on its diagonal
\end{description}

\item Expected values
\begin{description}
	\item[$\E {[x]} $] expected value of x
	\item[$\statecov_{ \fun(\state)\state}$] covariance of $\fun(\state)$ and $\state$
 	\item[$\mean_\state$] mean of $\state$
\end{description}

\item Other variables
\begin{description}
	\item[$I$] identity matrix
 	\item[$\jacobian$] matrix that defines the statistically linearized relationship between state and measurement, which is the Jacobian matrix in linear systems
 	\item[$\kalmangain$] Kalman gain
 	\item[$\innocov$] innovation covariance
	\item[$\mathbf{0}$] zero matrix
	\item[$\mathbf{1}$] matrix containing ones
\end{description}

\item Acronyms 
\begin{acronym}[SOKF2]
\acro{CKF}{Cubature Kalman Filter}
\acro{DD}{Divided Difference}
\acro{EKF}{Extended Kalman Filter}
\acro{EKF2}{Second Order Extended Kalman Filter}
\acro{EnKF}{Ensemble Kalman Filter}
\acro{GF}{Gaussian Filter}
\acro{GMF}{Gaussian Mixture Filter}
\acro{KF}{Kalman Filter}
\acro{KFE}{Kalman Filter Extension}
\acro{PDR}{Pedestrian Dead Reckoning}
\acro{QKF}{Gauss-Hermite Quadrature Kalman filter}
\acro{RBPF}{Rao-Blackwellized Particle Filter}
\acro{RUF}{Recursive Update Filter}
\acro{SLAM}{Simultaneous Localization and Mapping}
\acro{SOKF2}{Second Order Polynomial Kalman Filter}
\acro{S2KF}[S\textsuperscript{2}KF]{Smart Sampling Kalman Filter}
\acro{TDoA}{Time Difference of Arrival}
\acro{UKF}{Unscented Kalman Filter}
\acro{URUF}{Unscented Recursive Update Filter}
\end{acronym}
\end{itemize}

\section{Background}
\label{sec:background}
In discrete-time Bayesian filtering, the state of a dynamical system is estimated based on noisy measurements.  The state model describes how the $\statedim$-dimensional state $x$ is propagated in time. A state model can be expressed as
\begin{equation}
	x_t = \nonlinearstatetrans_{t}(\state_{\tm}, \statenoise_t), 
\end{equation}
where $\nonlinearstatetrans(\cdot)$ is the state transition function, $\state_{\tm}$ is the state at the previous time step, and $\statenoise_t$ is the state transition noise. The state's distribution is updated using measurements of the form
\begin{equation}
	y_t = \measfun_{t}(\state_{t}, \measnoise_t), 
\end{equation}
where $y_t$ is the realized measurement, $\measfun(\cdot)$ is the measurement function, and $\measnoise$ is the measurement noise.

To shorten notations we use the augmented state $ z =\smatr{\state\\ \noise} $, where applicable. In the augmented state $\noise$ is either state transition noise or measurement noise, depending on context. The state and measurement noises are assumed independent and white. We also assume that all the variables of interest are in the state. If the state or measurement noises are not white, they can be modeled using more  variables in the state.  \uus{schmidt}{In the Schmidt-Kalman filter, the additional noise bias states are not estimated, instead their effect on the covariance is approximated. However, the Schmidt-Kalman filter is suboptimal and we do not treat it in this paper. The interested reader may refer to \cite[p.282]{jazwinski}}. We omit the time index $t$ when its absence should not cause confusion.

In general, a \ac{KFE} can be divided into two parts:
\begin{enumerate}
\item Prediction:
\begin{align}
	\mean_{x_t}^- &= \mean_{\nonlinearstatetrans(z_{\tm})} \\
	\statecov_{x_t,x_t}^- &= \statecov_{{\nonlinearstatetrans(z_{\tm})}{\nonlinearstatetrans(z_{\tm})}},
\end{align}
where $\mean_{x_t}^-$ is the predicted mean computed using the state transition function and posterior of the previous time step and $\statecov_{x_t,x_t}^-$ is the predicted covariance.  
\item Update:
\begin{align}
	\meas_t^- &= \mean_{\measfun(z_t^-)} \\
	\innocov_t & =   P_{\measfun(z_t^-)\measfun(z_t^-)} \label{equ:innocov}\\
	\kalmangain_t &= P_{x_t^-\measfun(z_t^-)}\innocov_t ^{-1} \label{equ:Kalmangain} \\
	\mean_{x_t} &= \mean_{x_t}^- + K(\meas_t - \meas_t^-)  \\
	\statecov_{x_t,x_t} & = \statecov_{x_t,x_t}^- -K_tS_tK_t^T,
\end{align}
	where $z_t^-$ is the augmented state of predicted state and measurement noise, $\meas_t^-$ is the predicted mean of the measurement, $\innocov_t$ is the innovation covariance, $\kalmangain_t$ is the Kalman gain, $\mean_{x_t} $ is the updated mean of the state, and $\statecov_{x_t,x_t}$ is the updated state covariance. In \eqref{equ:Kalmangain}, $P_{x_t^-\measfun(z_t^-)}$ refers to the rows of $P_{z_t^-\measfun(z_t^-)}$ that correspond to the state variables. 
\end{enumerate}
The mean and covariance matrices associated to measurement and state transition functions can be written using expectations:
\uusi{brackets}{
\begin{align}
	\mu_{\fun(z)} & =\Ex{\fun(z)} \label{equ:E1}\\
	P_{\fun(z)\fun(z)} & =  \Ex{(\fun(z)-\mu_{\fun(z)})(\fun(z)-\mu_{\fun(z)})^T}  \label{equ:E2}\\
	P_{z\fun(z)} & = \Ex{(z- \mu_z  ) \left( \fun(z)-\mu_{\fun(z)} \right)^T}\label{equ:E3}.
\end{align}}
When a function is linear, that is, of the form
\begin{equation}
	\fun(z) = \jacobian z,
\end{equation}
 the expectations have analytic form:
\begin{align}
	\mu_{\fun(z)} & = J\mu_z \\
	P_{\fun(z)\fun(z)} & =  JP_{zz}J^T \label{equ:JPJ}\\
	P_{z\fun(z)} & = P_{zz}J^T \label{equ:PPJ}
\end{align}
and the algorithm is the \ac{KF}. 

There are extensions that approximate the expectations \eqref{equ:E1}--\eqref{equ:E3} using analytic differentiation (or integration) of the functions. For example, the \ac{EKF} uses the first order Taylor expansion and the \ac{EKF2} does the linearization based on the second order Taylor expansion \cite{jazwinski}. There are also algorithms that use numerical differentiation to approximate  \eqref{equ:E1}--\eqref{equ:E3}. In \ac{DD} filters~\cite{norgaard} the computation is based on numerical differentiation of the functions to obtain a linear model. \ac{DD} filters use $2\statedim+1$ function evaluations. The \ac{SOKF2}~\cite{ITO} uses $ \frac{ 1}{2}n^2 + \frac{3}{2}\statedim + 1$ points to fit a second order polynomial to the function and then computes the analytic covariance matrices for the polynomial.

One very commonly used and large group of \acp{KFE} algorithms approximate expectations \eqref{equ:E1}--\eqref{equ:E3} as:
\begin{align}
	\mu_{g(z)} &\approx  \sum \sigmapw_i^s g(\chi_i) \label{equ:UKF1}\\
	P_{g(z)g(z)} & \approx  \sum \sigmapw_i^c (g(\chi_i) - \mu_{g(z)}) (g(\chi_i) - \mu_{g(z)})^T \label{equ:UKF2}\\
	P_{zg(z)} &\approx  \sum \sigmapw_i^c (\chi_i - z) (g(\chi_i) - \mu_g(z))^T, \label{equ:UKF3}
\end{align}
where $\chi_i$ are so-called sigma-points that are chosen according to the prior distribution and $w_i^s$ and $w_i^c$ are associated weights. Examples of this kind of filters are \ac{UKF}~\cite{UKF}, different \acp{CKF}~\cite{cubature}, and  \ac{QKF}~\cite{ITO}.

The selection and number of sigma-points and weights depend on the algorithm used. The \ac{UKF} is  usually used with $2\statedim +1$ sigma-points. The \ac{GF} uses $k\statedim+1$ sigma-points, where $k > 1$ is an integer parameter \cite{huber2008gaussian}. \acp{CKF} are developed for different orders and they use $\mathcal{O}(\statedim^o)$ sigma-points, where $o$ is the order of the cubature rule. The number of sigma-points in \acp{QKF} increases exponentially, as the number of sigma-points is $\alpha^\statedim$, where $\alpha > 1$~\cite{ITO}. There are also algorithms that allow an arbitrary number of points, for example \ac{S2KF}~\cite{S2KF}, which uses at least $\statedim+1$ sigma-points.

Algorithms that use \eqref{equ:UKF1}--\eqref{equ:UKF3} do not compute  $\jacobian$ explicitly. In some optimizations  $\jacobian$ is required and can be computed using statistical linearization~\cite{sarkka2013bayesian}
\begin{equation}
	 \jacobian  = P_{\fun(z)z} P_{zz}^{-1}. \label{equ:virtJac}
\end{equation}
%This matrix defines the relationship between $P_{zz}$ and $P_{\fun(z)z}$.
\uusi{jaccom}{For nonlinear systems, this cannot be substituted into \eqref{equ:JPJ} to obtain $P_{\fun(z)\fun(z)}$.} Because the computation of~\eqref{equ:virtJac} requires solving a set of linear equations it should be avoided when the dimension of $z$ is large.

The computational requirements of the algorithms can be reduced in several ways. One way is to compute the expected values numerically only for state variables that are transformed by a nonlinear function and compute the expectations for linear parts analytically \cite{7383842}. The improving of implementation efficiency using the linear \ac{KF} update for the linear state variables is more familiar in particle filtering. Specifically \ac{RBPF} solves the estimates of conditionally linear variables using a \ac{KF} and rest of the variables using particles \cite{Doucet:2000:RPF:647234.720075}. The dimension of the output of the nonlinear part can also be reduced to reduce the computational burden. In \eqref{equ:UKF2} the $m\times m$ covariance matrix is updated for every sigma point. Thus, halving the dimension $m$ of $\fun(\cdot)$  reduces the operations applied to the covariance matrix \uus{byfour}{by a factor of four}. Such update optimizations are considered in Section~\ref{sec:optimizations}.

It is also possible to compute the Kalman gain \eqref{equ:Kalmangain} faster by exploiting the structure of the innovation covariance \eqref{equ:innocov} as shown in Section~\ref{sec:inno}.

When the state can be divided into decoupled blocks\uus{inc}{, the} blocks can be updated independently and the global estimate can be computed only when needed. The updates of the blocks can be made \uus{inat}{at} different rates. These optimizations are presented in Section~\ref{sec:multiprocess}.

\section{Optimizations Based on the Linearity in Nonlinear functions}

\label{sec:optimizations}

\subsection{Partially Linear Functions}
\label{sec:partlin}
In~\cite{1257248}, algorithms for treating different setups of \ac{UKF} were presented. The algorithms considered different implementations when the state transition model or measurement model has one of the following three forms
\begin{align}
	\fun(\state,\noise)&=\Mfun\state+\noise \\ 
	\fun(\state,\noise) & =\fun_1(\state)+\noise \label{equ:additivenoise}\\
	\fun(\state,\noise) & =\fun_2(\state,\noise),
\end{align}
where noise $\noise$ is assumed independent of the state. The algorithms improve \uus{the}{in} computation efficiency when either or both of the state and measurement models belong to the first or second group. If both are in the first group the updates can be computed exactly using the \ac{KF}. If the function belongs to the second group then \eqref{equ:E1}--\eqref{equ:E3} 
become
\begin{align}
	\mu_{\fun(z)} & = \E \left[\fun(x)\right] + \mean_\noise \label{equ:U1}\\
	P_{\fun(z)\fun(z)} & =  \E \left[ (\fun(x)-\mu_{\fun(x)})(\fun(x)-\mu_{\fun(x)})^T\right]  + P_{\noise\noise} \label{equ:U2}\\
	P_{x\fun(z)} & = \E \left[ (x- \mu_x  ) \left( \fun(x)-\mu_{\fun(x)} \right)^T \right] \label{equ:U3}
\end{align}
and the approximation can be computed for the dimension of the state instead of the dimension of the augmented state. This optimization is widely used among different filters and is often considered to be the standard model. 

In~\cite{1660584} the function is split into a nonlinear $\fun_1(\cdot)$ part that depends only on a part of the  state $z_n$  and linear parts  
\begin{equation}
	\measfun(z) = \matr{ \fun_1(z_n ) \\ H_1 z}. \label{equ:lnonl}
\end{equation}
The \ac{UKF} is used for computing the expected values of the nonlinear part and the correlation between nonlinear and linear parts.

We generalize the above models to functions of form:
\begin{equation}
	g(z) = A g_n\left(Tz \right) + Hz, \label{equ:fullequ}
\end{equation}
where $T$ has full row rank.  To reduce the required resources in computation $T$  is chosen to have the minimal number of rows and $g(\cdot)$ to have the minimal number of elements. Expectations \eqref{equ:E1}--\eqref{equ:E3} are computed for $Tz$, which should have a smaller dimension than $z$. 

For computing the update, an algorithm that approximates expectations \eqref{equ:E1}--\eqref{equ:E3} is required. These are usually computed in the update stage of a \ac{KFE}. The cross correlation matrix $P_{zg(z)}$ \eqref{equ:UKF3} is not needed in the normal state propagation, but the algorithm from the update stage can be used for this.

The transformed augmented state is denoted
\begin{equation}
\tilde z = Tz \sim \N(\mu_{\tilde z}=T\mu_z,P_{\tilde z\tilde z}= TP_{zz} T^T   ). \label{equ:transt}
\end{equation}
When expectations 
$\mean_{g_n(\tilde z)}$ \eqref{equ:E1}, $P_{g_n(\tilde z)g_n(\tilde z)}$ \eqref{equ:E2}, and $P_{\tilde zg_n(\tilde z)}$  
\eqref{equ:E3} are known, the expectations for \eqref{equ:fullequ} are \uusi{fixc}{
\begin{align}
	\mean_{g(z)} & = A \mean_{g_n(\tilde z)} + H\mean_z \label{equ:mom1} \\ 
	P_{g(z)g(z)} &= \matr{A & H} \matr{ P_{g_n(\tilde z)g_n(\tilde z)} & P_{zg_n(\tilde z)}^T \\ P_{zg_n(\tilde z)} & P_{zz} } \matr{ A^T \\  H^T } \label{equ:isoP}\\
	P_{z g(z)} &= \matr{ P_{zg_n(\tilde z)} & P_{zz}  }\matr{A^T \\ H^T}, \label{equ:mom3}
\end{align}}
where
\uusi{gn}{
\begin{align}
	P_{zg_n(\tilde z)} & = P_{zz}T^T\left(T P_{zz} T^T \right)^{-1} P_{\tilde zg_n(\tilde z)}  . \label{equ:replace}
\end{align}}
This is based on the fact that the cross term $P_{\tilde zg_n(\tilde z)}$ describes the linear dependence of function $g_n(\tilde z)$ and $\tilde z$ as in \eqref{equ:virtJac}: 
\begin{equation}
P_{\tilde zg_n(\tilde z)} = P_{\tilde z\tilde z}J^T  = TP_{zz}T^TJ^T \label{equ:Pz}.
\end{equation}
and the term $P_{zg_n(\tilde z)}$ is
\begin{equation}
P_{ zg_n(\tilde z)} = P_{zz} T^T J^T. \label{equ:pzg}
\end{equation}
Solving $J$ from \eqref{equ:Pz} and substituting it into \eqref{equ:pzg} we get \eqref{equ:replace}.

In the update, the matrix with cross terms is required for state variables only. This can be extracted by
\begin{equation}
	P_{x\measfun(z)} = \matr{I & 0}P_{z\measfun(z)}  \label{equ:pxh}
\end{equation}
when the state variables appear first in the augmented state. Naturally, in a computationally efficient code~\eqref{equ:pxh} is done with indices, not matrix multiplication. 

\begin{algorithm}[t]
\caption{State transition using a partially linear measurement}
\label{algo:partiallinearpred}
%\scriptsize
\algoline

%\SetAlgoNoLine
\KwIn{
	  $x_{\tm} \sim \N(\mean_{x_{\tm}}, P_{x_{\tm}x_{\tm}})$ \tcp{State estimate from previous time step}
          $\transnoise \sim \N(\mean_{\transnoise},P_{\transnoise \transnoise}  )$,  \tcp{State transition noise}
	 $P_{x_{\tm} \transnoise}$ \tcp{Cross covariance between state and state transition noise}
	 $\statetransfun(\state,\transnoise)=\statetransfun(z)=A g\left(Tz \right) + Hz$,  \tcp{State transition function of given form}
}
  \KwOut{
	$\state \sim \N( \mean_{\state^-}, P_{x^-x^-} )$ \tcp{Propagated state} 
}
$\mean_{z} = \smatr{ \mean_{x_{\tm}} \\ \mean_{\transnoise}}$ \tcp{Augmented mean}
$P_{zz} = \smatr{ P_{x_{\tm}x_{\tm}} & P_{x_{\tm} \transnoise} \\ P_{x_{\tm} \transnoise}^T & P_{\transnoise \transnoise}}$ \tcp{Augmented covariance}

$\mean_{\tilde z} = T\mean_z$ \tcp{Transformed mean}
 $P_{\tilde z\tilde z} = TP_{zz}T^T$ \tcp{Transformed covariance}
Compute $\mean_{g(\tilde z)}$,$P_{g(\tilde z)g(\tilde z)}$, and $P_{\tilde zg(\tilde z)}$ using a \ac{KFE} \\
$P_{zg(\tilde z)}  = P_{zz}T^T\left(T P_{zz} T^T \right)^{-1} P_{\tilde zg(\tilde z)}$ \\
$\mean_{\state^-}  = A \mean_{g(Tz)} + H\mean_z$  \tcp{Predicted mean of state}
$P_{\state^-\state^-} = \smatr{A & H} \smatr{ P_{g(\tilde z)g(\tilde z)} & P_{zg(\tilde z)}^T \\ P_{zg(\tilde z)} & P_{zz} } \smatr{ A^T \\  H^T }$  \tcp{Predicted covariance}
\algoline

\end{algorithm}

\begin{algorithm}[tbp!]
\caption{Update using a partially linear measurement}
\label{algo:partiallinearup}
%\scriptsize
\algoline

\KwIn{
	  $x^- \sim \N(\mean_{x^-}, P_{x^-x^-})$ \tcp{Prior state}
          $\measnoise \sim \N(\mean_{\measnoise},P_{\measnoise \measnoise}  )$,  \tcp{Measurement noise}}
        	 $P_{x^- \measnoise}$ \tcp{Cross covariance between state and measurement noise}
          $\measfun(\state,\measnoise)=\measfun(z)=A g\left(Tz \right) + Hz$,  \tcp{Measurement function of given form}
\KwOut{
	$\state \sim \N( \mean_\state, P_{xx} )$ \tcp{Posterior estimate} 
}
$\mean_z = \smatr{\mean_{x^-} \\ \mean_{\measnoise}}$  \tcp{Mean of the augmented state} 
$P_{zz} = \smatr{P_{x^-x^-} & P_{x^-\measnoise} \\ P_{x^-\measnoise}^T & P_{\measnoise\measnoise}}$ \tcp{Covariance matrix of the augmented state}
$\mean_{\tilde z} = T\mean_z$ \tcp{Transformed mean}
 $P_{\tilde z\tilde z} = TP_{zz}T^T$ \tcp{Transformed covariance}
Compute $\mean_{g(\tilde z)}$,$P_{g(\tilde z)g(\tilde z)}$, and $P_{\tilde zg(\tilde z)}$ using a \ac{KFE} \\
$P_{zg(\tilde z)}  = P_{zz}T^T\left(T P_{zz} T^T \right)^{-1} P_{\tilde zg(\tilde z)}$ \\
$\mean_{\measfun(z)}  = A \mean_{g(Tz)} + H\mean_z$  \tcp{Predicted mean of measurement}
$P_{\measfun(z)\measfun(z)} = \smatr{A & H} \smatr{ P_{g(\tilde z)g(\tilde z)} & P_{zg(\tilde z)}^T \\ P_{zg(\tilde z)} & P_{zz} } \smatr{ A^T \\  H^T }$  \tcp{Innovation covariance}
$P_{x \measfun(z)} = \smatr{ P_{zg(\tilde z)} & P_{zz}  }_{[1:\statedim,:]}\smatr{A^T \\ H^T}$ \tcp{State-measurement cross correlation} 
$K=P_{x \measfun(z)} P_{\measfun(z)\measfun(z)}^{-1}$ \tcp{Kalman Gain}
$\mean_\state = \mean_\state^- + K (y-\mean_{\measfun(z)})$ \tcp{Posterior mean}
$P_{\state\state} = P_{x^-x^-} - KP_{\measfun(z)\measfun(z)} K^T$ \tcp{Posterior covariance}
\algoline
\end{algorithm}

The algorithm for state transition is given in Algorithm~\ref{algo:partiallinearpred} and an algorithm for state update is given in  Algorithm~\ref{algo:partiallinearup}.  Use of these algorithms is beneficial when the dimension of $\tilde z$ is smaller than $z$ and the matrix inverse in \eqref{equ:replace} is applied in a small dimension. Algorithms~\ref{algo:partiallinearpred} and~\ref{algo:partiallinearup} are given in a general form and for applications they can be optimized further by considering the structures of matrices. Examples of exploiting the partially linear state are given in sections~\ref{sec:fourier},\ref{sec:pdr}, and~\ref{sec:tdoa}.

\subsection{Conditionally Linear Measurements}
\label{sec:condlin}
In~\cite{4217985}, a situation where a function can be divided into nonlinear $z_n$ and conditionally linear $z_l$ parts
\begin{equation}
	\fun(z) = \fun_n(z_n )+ \Mfun_n( z_n )z_l \label{equ:condl}
\end{equation}
is considered. In the original article, the distribution of this nonlinear function is approximated using a modification of the \ac{UKF}. The number of sigma-points depends on the dimension of  $z_n$ instead of the dimension of the full state.

This algorithm is converted for use with the \ac{GF} in~\cite{5203794}.  Here we present the algorithm in a general form that allows it to be used with any \ac{KFE} that uses weighted points to compute the expectations, as in \eqref{equ:U1}--\eqref{equ:U3}, although it cannot be used with other types of filters e.g.\ \ac{DD} or \ac{SOKF2}.

In the algorithm the sigma-points are computed for the nonlinear part only and then used for computing the conditional probabilities of the conditionally linear part:
\begin{align} 
	\mu_{z_l | \sigmap_i}  & = \mu_{z_l} +  P_{z_lz_n}P^{-1}_{z_nz_n}(\sigmap_i - \mu_{z_n}) \label{equ:condmu}\\
	P_{z_lz_l | \sigmap_i } & = P_{z_lz_l} - P_{z_lz_n}P^{-1}_{z_nz_n}P_{z_nz_l}. \label{equ:condP}
\end{align}
The matrices in  (\ref{equ:condmu}--\ref{equ:condP}) are independent of the sigma-point $\sigmap_i$. Thus, $P_{z_lz_n}P^{-1}_{z_nz_n}$ and $P_{z_lz_l | \sigmap_i }$ need to be computed only once. According to~\cite{4217985},  the expectations for a function of form \eqref{equ:condl} can be approximated as 
\begin{align}
{\mathcal Y}_i = &  g_n (\sigmap_i)    +\Mfun_n (\sigmap_i)  \mu_{z_l | \sigmap_i} \label{equ:condl1}\\
\mu_{\fun(z)}  = &\sum w_i {\mathcal Y}_i \ \\ 
P_{\fun(z)\fun(z)}  =&  \sum w_i \left(\left(  {\mathcal Y}_i - \mu_{\fun(z)}  \right)\left({\mathcal Y}_i  - \mu_{\fun(z)}   \right)^T \right. \notag \\ & \left. \hspace{1.5cm }+  \Mfun_n(\sigmap_i)  P_{z_l z_l | \sigmap_i}  \Mfun (\sigmap_i)^T\right) \\
P_{z\fun(z)} = &\sum w_i\left( \left( \smatr{\sigmap_i \\ \mu_{z_l | \sigmap_i}} - \mu_{z} \right)\left( {\mathcal Y}_i  - \mu_{\fun(z)} \right)^T  \right.\notag \\ & \left. \hspace{1.5cm }  + \smatr{0 \\P_{z_l z_l | z_n}\Mfun_n(\sigmap_i)^T }\right).  \label{equ:condl3}
\end{align}
These formulas can be used to compute the expectations for the nonlinear part $g(\tilde z)$ in algorithms~\ref{algo:partiallinearpred} and~\ref{algo:partiallinearup}, if there are conditionally linear state variables. This is done in the example in sections~\ref{sec:pdr} and~\ref{sec:lithium}.

\section{Optimizations Related to the Inverse of the Innovation Covariance}
\label{sec:inno}
\subsection{Block Diagonal Measurement Covariance}
\label{sec:Innopti}
The formula for the Kalman gain \eqref{equ:Kalmangain} contains the inverse of the innovation covariance~$\innocov$. When the dimension of the innovation covariance (i.e.\ the number of measurements) is large the computation of the inverse or solving the set of linear equations is a computationally expensive operation.

When measurements are linear and the measurement covariance is diagonal, the Kalman update can be applied one measurement element at a time and the partially updated state used as a prior for the next measurement \cite{stengel1986stochastic}. This kind of update can be generalized also to block diagonal measurement covariances and the update can be done by applying one block at a time. This reduces the computation time when the state dimension is small, the measurement dimension is large, and the measurements are independent.

When some measurements are independent and these different measurements contain different state terms, updating the state using only a part of measurements at once can be effectively combined with the algorithms presented in Section~\ref{sec:optimizations}. The block update does not change the estimate when the measurement model is linear, but when the model is nonlinear the linearization is redone in the partially updated state. This may alter the output of the algorithm. In~\cite{PUKF,Raitoharju2016} application of measurements one element at a time is used to improve the estimation accuracy by applying the most linear measurements first. The algorithm for updating by blocks is given in Algorithm~\ref{algo:blocup}.
\begin{algorithm}[bp]
\caption{Block Update \ac{KF}}
\label{algo:blocup}
%\scriptsize
\algoline

\KwIn{
	  $x_0 \sim \N(\mean_{x,0}, P_{xx,0})$ \tcp{Prior state}
          $\measnoise_i \sim \N(\mean_{\measnoise_i},P_{\measnoise_i \measnoise_i}  )$, $1\leq i \leq n$ \tcp{Measurement noises}}
          $\measfun_i(\state,\measnoise_i)$, $1\leq i \leq n$ \tcp{Measurement functions}
\KwOut{
	$x_n \sim \N( \mean_{x,n}, P_{xx,n} )$ \tcp{Posterior estimate} 
}
\For{i=1 to n}
{
	$\meas_{i}^- = \mean_{\measfun_i( x_{i-1}, \measnoise_i )}$ \\
	$\innocov_{i}  =   P_{\measfun_i(z_{t,i}^-)\measfun(x_{i-1}, \measnoise_i)}$ \\
	$\kalmangain_{i} = P_{x\measfun_i(z_{t,i}^-)}\innocov_{i}^{-1}$ \\
	$\mean_{x,i} = \mean_{x,i-1}  + K( \meas_{i} - \meas_{i}^-)$  \\
	$P_{xx,i}  = P_{xx,i-1} -K_{i}S_{i}K_{i}^T$
}
\algoline
\end{algorithm}

\subsection{Applying the Matrix Inversion Lemma to Innovation Covariance}
\label{sec:innopti2}
When the measurements are not independent the block update formula cannot be used. In some situations the innovation covariance can be written in the form
\begin{equation}
	S = P_s + UP_vU^T, \label{equ:innoSSS}
\end{equation}
where $P_s$ is easy to invert, $P_v$ has a small dimension and $U$ is a transformation matrix.

Using the matrix inversion lemma \cite[p. 62]{stengel1986stochastic} the inverse of \eqref{equ:innoSSS} is
\begin{equation}
	S^{-1} = P_s^{-1} - P_s^{-1} U\left( P_v^{-1} + UP_s^{-1}U^T  \right)^{-1} U^T P_s^{-1}. \label{equ:woodbury}
\end{equation}
This formula is worth using if the inverse of $P_s$ is easy to compute or can be computed offline and the dimension of $P_v$ is smaller than the dimension of  $P_s$.  An example of its use is given in Section~\ref{sec:tdoa}.

\section{Optimization Based on Dividing the State into Individual Estimation Processes}
\label{sec:multiprocess}
In this section, we study optimizations that can be applied when the state can be divided into separate subprocesses. The situation where only part of state variables occur in the measurements and the part of the state that is not in measurements does not change in time has been considered in \cite{davison1998mobile,938382}. In that situation the estimation process can be done at each time step only for the part of the state that has been observed and then the whole state is updated only occasionally. 

Such algorithms are developed further in \cite{guivant_2017}. The algorithm given in \cite{guivant_2017}  assumes that the state can be divided into substates that can be updated individually.  The division can be done when the measurement and state transition models in a time interval from $t_0$ to $t_1$ are decoupled between the substates.  State blocks $x_{\part{1}}, x_{\part{2}},\ldots$ are considered to be decoupled if the  state transition and measurement models can be expressed as
\begin{align}
	\statefun_t(x,\varepsilon) & = \matr{ \statefun_{\part{1},t}(x_{\part{1},t-1},\statenoise_{\part{1},t}) \\ \statefun_{\part{2},t}(x_{\part{2},t-1},\statenoise_{\part{2},t})  \\ \vdots} \label{equ:cond1}\\
	\measfun_t(x,\varepsilon) & = \matr{ \measfun_{\part{1},t}(x_{\part{1},t},\measnoise_{\part{1},t}) \\  \measfun_{\part{2},t}(x_{\part{2},t},\measnoise_{\part{2},t}) \\ \vdots} \label{equ:cond2}, 
\end{align}
where noise terms $\statenoise$ and $\measnoise$ are independent.  The purpose of this optimization is to allow to track state variables that are decoupled from time index $t_0$ to time index $t_1$.  Note that there is no requirement for blocks $x_{\part{1}},x_{\part{2}}, \ldots$ to be independent at $t_0$. If the blocks were independent at $t_0$ and models could be expressed using \eqref{equ:cond1} and \eqref{equ:cond2} in all time instances, then the system could be solved using a set of independent \acp{KFE}.

A similar idea for splitting the state into multiple blocks is presented in \cite{6302204,7592912} under the name of multiple quadrature Kalman filtering. In multiple quadrature Kalman filtering, there is no strict requirement for blocks to be decoupled, which makes the algorithm applicable to a larger set of problems, but leads to additional approximations, so we follow \cite{guivant_2017}.

To update block $x_{\part{a}}$ from time $t_0$ to $t_1$ a new Gaussian variable $\bar{x}$, which is partitioned into two parts $x$ and $\hat{x}$, is introduced:
\begin{align}
	\bar{x}_{\part{a} ,t_0}=\matr{x_{\part{a},t_0} \\ \hat x_{\part{a},t_0}} &
	\sim \N\left( \matr{ \mean_{x_{\part{a},t_0}} \\  \mean_{\hat x_{ \part{a} ,t_0}} },   \matr{ P_{x_{\part{a},t_0} x_{\part{a},t_0}} & P_{x_{\part{a},t_0} {\hat x_{\part{a},t_0}}} \\ P_{{\hat x}_{\part{a},t_0}x_{\part{a},t_0}} & P_{{\hat x}_{\part{a},t_0} {\hat x}_{\part{a},t_0} }  
	}  \right) \notag \\ &= \N\left(\matr{ {\mean_x}_{\part{a},t_0} \\  {\mean_x}_{\part{a},t_0} } , \matr{  \statecov_{x_{\part{a},t_0}x_{\part{a},t_0}} & \statecov_{x_{\part{a},t_0}x_{\part{a},t_0}} \\ \statecov_{x_{\part{a},t_0}x_{\part{a},t_0}}  & \statecov_{x_{\part{a},t_0}x_{\part{a},t_0}}  }  \right).
\end{align}
In the estimation of this, initially degenerate, variable from time $t_0$ to $t_1$ $x_{\part{a}}$ is updated and propagated normally and $\hat{x}_{\part{a}}$ changes only through its dependence on part $x_{\part{a}}$.

In the prediction, state elements belonging to $x_{\part{a}}$ are transformed and $\hat{x}_{\part{a}}$ remains static:
\begin{equation}
	\matr{x_{\part{a},t} \\ \hat x_{\part{a},t}}  = \matr{ f_{\part{a},t}(x_{\part{a},t-1}, \statenoise_{\part{a},t}) \\ \hat x_{\part{a},t-1}}. \label{equ:blockprob}
\end{equation}
The measurement function is applied only to the first elements
\begin{equation}
	h_{\part{a},t}\left( \matr{x_{\part{a},t} \\ \hat x_{\part{a},t}}, \measnoise_{\part{a},t} \right)	= h\left( x_{\part{a},t},\measnoise_{\part{a},t} \right). \label{equ:blockmeas}
\end{equation}
These updates can be done with any suitable \ac{KF} or other estimation method. The division of the state into blocks allows also updating blocks at different rates.

 The number of elements in the covariance matrix of each block is $(2 \statedim_\text{block})^2$, thus if a 100-dimensional state is divided into 100 decoupled blocks, the updates of blocks change 400 values in covariance matrices, while the update of the full state would change 10000 values. The estimates for each block contain only information about the measurements and state transition of that block, and not information that comes from other blocks through the dependencies of the blocks at time~$t_0$. 

At time $t_1$ the information of the different blocks is merged in a global update that is done in two phases: First a virtual update is constructed from all blocks and applied to the full state estimate at time index $t_0$. Then a virtual state prediction is  applied to the virtually updated full state. This order is the opposite of the usual Kalman filter update where prediction is usually done first and the update is done after that. The virtual update fuses the individual blocks so that the correlations that existed between the blocks at time $t_0$ are taken into account. To denote state parameters that have been updated with virtual update we use superscript $v$; parameters for which both the virtual update and prediction has been applied  are denoted with superscript $+$.

\subsection{Virtual update}
The virtual update \uus{virtual}{updates the state with a Kalman update that uses virtual observations $y^v$, measurement matrix $J^v$, and measurement covariance $R^v$ \cite{guivant_2017}. The parameters for a virtual update are}
\begin{align}
	   y^v_{t_1} &=  \matr{ y^v_{\part{1},t_1} \\ y^v_{\part{2},t_1} \\ y^v_{\part{3},t_1} \\ \vdots  }\\
	   J^v_{t_1} & =I \\
	R^v_{t_1} & = \matr{R^v_{\part{1},t_1} & \mathbf{0} & \ldots & \\ 
	              \mathbf{0} & R^v_{\part{2},t_1} & \mathbf{0} & \ddots \\ 
	              \vdots & \mathbf{0}  & R^v_{\part{3},t_1} & \ddots \\ 
	              & \ddots & \ddots &\ddots}, \label{equ:RRR}
\end{align}
where components have the form
\begin{align}
	y^v_{\part{a},t_1}  = & \mu_{x_{\part{a}},t_0} \notag  \\ 
	                       & +  P_{x_{\part{a}}x_{\part{a}},t_0} \left( P_{x_{\part{a}}x_{\part{a}},t_0} - P_{\hat{x}_{\part{a}}\hat{x}_{\part{a}},t_1}\right)^{-1}\left( {\hat\mu}_{{\part{a}},t_1}- \mu_{x_{\part{a}},t_0} \right) \label{equ:muv}\\
	R^v_{\part{a},t_1} &= P_{x_{\part{a}}x_{\part{a}},t_0} \left( P_{x_{\part{a}}x_{\part{a}},t_0} - P_{\hat{x}_{\part{a}}\hat{x}_{\part{a}},t_1}\right)^{-1}P_{x_{\part{a}}x_{\part{a}},t_0} \notag \\ & - P_{x_{\part{a}}x_{\part{a}},t_0}. \label{equ:Rv}
\end{align}
These both require the inversion of $\left( P_{x_{\part{a}}x_{\part{a}},t_0} - P_{\hat{x}_{\part{a}}\hat{x}_{\part{a}},t_1}\right)$ which may be singular. One example of it being singular is the situation of not having any measurements considering the current block. In \cite{guivant_2017}, the singular situation is handled by using singular value decomposition. 

In the following, we give a new formulation that does not require the singular value decomposition and is equivalent to the update presented in \cite{guivant_2017}. The formulation is derived in the Appendix. In the new formulation, the ''posterior'' parameters after the virtual update are
\allowdisplaybreaks[0]
\begin{align}
\mu_{x,t_1}^v = &\mu_{x,t_0} + \left(  \PhI - \PdI +   \PmI  \right)^{-1} \notag \\ & \cdot \left( \PhI  \hat{\mu}_{x,t_1}  -\PmI  \mu_{x,t_0} \right)  \label{equ:unobstart}  \allowdisplaybreaks \\
P_{xx,t_1}^v = &\left(  \PhI - \PdI +   \PmI  \right)^{-1}, \label{equ:covup}
\end{align}
where $P_{\hat{x}\hat{x},t_1}$ and $P^{D}_{\hat{x}\hat{x},t_0}$ are defined as
\begin{equation}
	\Ph = \matr{ P_{\hat{x}_{\part{1} }\hat{x}_{\part{1}},t_1}  & \mathbf{0} & \ldots  & \\
	                                                \mathbf{0} &  P_{\hat{x}_{\part{2} }\hat{x}_{\part{2}},t_1}  & \mathbf{0} & \ddots \\
	                                                \vdots & \mathbf{0} & P_{\hat{x}_{\part{3} }\hat{x}_{\part{3}},t_1}  & \ddots \\
	                                                 & \ddots & \ddots & \ddots} 
\end{equation}
\begin{equation}
	\Pd = \matr{ P_{{x}_{\part{1} }{x}_{\part{1}},t_0}  & \mathbf{0} & \ldots  & \\
	                                                \mathbf{0} &  P_{{x}_{\part{2} }{x}_{\part{2}},t_0}  & \mathbf{0} & \ddots \\
	                                                \vdots & \mathbf{0} & P_{{x}_{\part{3} }{x}_{\part{3}},t_0}  & \ddots \\
	                                                 & \ddots & \ddots & \ddots}.
\end{equation}
Because these matrices are block diagonal, their inverses can be computed blockwise.

This formulation does not require inversion of singular matrices, provided that
\begin{itemize}
	\item Prior $P_{xx,t_0}$ has full rank
	\item Posterior $P_{xx,t_1}$ has full rank
\end{itemize}

\subsection{Virtual state propagation}
After all blocks are updated as presented in the previous subsection the state is propagated with a virtual state propagation to obtain the global posterior. The state propagation model is linear
\begin{equation}
	x = F^v(x^v - \hat\mu) + \mu + \varepsilon^v,
\end{equation}
where $\varepsilon^v$ has zero mean and covariance $Q^v$. Thus the posterior mean and covariance are \begin{align}
	\mean_{\state,t_1}^+  &= F^v\mu^{v} - F^v\hat\mu +\mu \label{equ:blockposteriormean} \\ 
         \statecov_{\state,\state,t_1}^+ &= F^vP^v{F^v}^T+Q^v \label{equ:blockposteriorcov} 	
\end{align}

The state propagation parameters have the following form
\begin{align}
	\mu &= \matr{ \mu_{x_{\part{1}},t_1}^T & \mu_{x_{\part{2}},t_1}^T & \ldots }^T \\
	\hat{\mu} &= \matr{ \hat{\mu}_{x_{\part{1}},t_1}^T & \hat{\mu}_{x_{\part{2}},t_1}^T & \ldots }^T \\
	F^v &= \matr{F_{\part{1}} & \mathbf{0} & \ldots \\
	                \mathbf{0} & F_{\part{2}} & \mathbf{0} & \ddots \\
			\vdots & \mathbf{0} &  F_{\part{3}} & \ddots \\ 
			          & \ddots & \ddots & \ddots } \\
	Q^v &= \matr{Q_{\part{1}} & \mathbf{0} & \ldots \\
	                \mathbf{0} & Q_{\part{2}} & \mathbf{0} & \ddots \\
			\vdots & \mathbf{0} &  Q_{\part{3}} & \ddots \\ 
			          & \ddots & \ddots & \ddots },
			\end{align}
where the matrix blocks are 
\begin{align}
	F_{\part{a}}^v &= P_{x_{\part{a}}\hat{x}_{\part{a}},t_1} P_{\hat{x}_{\part{a}}\hat{x}_{\part{a}},t_1}^{-1} \\
	Q_{\part{a}}^v &= P_{x_{\part{a}}x_{\part{a}},t_1}-  P_{x_{\part{a}}\hat{x}_{\part{a}},t_1} P_{\hat{x}_{\part{a}}\hat{x}_{\part{a}},t_1}^{-1} P_{\hat{x}_{\part{a}}x_{\part{a}},t_1}. \label{equ:unobend}
\end{align}
The full algorithm for filtering a model whose state has decoupled blocks is given in Algorithm~\ref{algo:fullblock}.
\begin{algorithm}[tbp]
\caption{\ac{KF} with decoupled blocks}
\label{algo:fullblock}
%\scriptsize
\algoline

\KwIn{
	  $x_{t_0} \sim \N(\mean_{x,0}, P_{xx,0})$ \tcp{Prior state that can be divided into decoupled blocks}
          $\statetransfun_{t,{\part{a}}}(\state_{\part{a}},\statenoise_{t,{\part{a}}})$ \tcp{State transition functions for blocks for each time step}
          $\measfun_{t,{\part{a}}}(\state_{\part{a}},\measnoise_{t,{\part{a}}})$ \tcp{Measurement functions for blocks for each time step}
		\tcp{Note that each block may have different number of state transition and measurement functions between $t_0$ and $t_1$}
          }
\KwOut{
	$x \sim \N( \mean^+_{x,t_1}, P^+_{xx,t_1} )$ \tcp{Posterior estimate} 
}
\For{a=1 to n}
{
	${\overline \mu}_{x_{\part{a}}} = \matr{ {\mean_x}_{\part{a},t_0} \\  {\mean_x}_{\part{a},t_0} }$ \tcp{Initialize block mean}
	${\overline \statecov}_{x_{\part{a}}x_{\part{a}}}  = \matr{  \statecov_{x_{\part{a},t_0}x_{\part{a},t_0}} & \statecov_{x_{\part{a},t_0}x_{\part{a},t_0}} \\ \statecov_{x_{\part{a},t_0}x_{\part{a},t_0}}  & \statecov_{x_{\part{a},t_0}x_{\part{a},t_0}} }$ \tcp{Initialize block covariance}
}

\For{$a=1$ to $n$}
{
	\For{$t_{\part{a}}=t_{0}$ to $t_1$}
{
	Update ${\overline \mu}_{x_{\part{a}}}$ and ${\overline \statecov}_{x_{\part{a}}x_{\part{a}}}$ with the state propagation and measurement models at time  $t_{\part{a}}$ using a suitable filter and \eqref{equ:blockprob}--\eqref{equ:blockmeas}.
}
}
\tcp{Update global state after each block is updated to time $t_1$}
Compute virtual prediction mean $\mean^v_{x,t_1}$ with \eqref{equ:unobstart} and covariance $P^v_{xx,t_1}$with \eqref{equ:covup}.\\
Compute virtual prediction mean $\mean^+_{x,t_1}$ with \eqref{equ:blockposteriormean} and covariance $P^+_{xx,t_1}$ with \eqref{equ:blockposteriorcov}.
\algoline
\end{algorithm}

\subsection{Static blocks}
In this section, we consider a common special case that allows to optimize the formulas in the  previous sections further. This corresponds to the algorithm presented in \cite{LefebvreBS05}.

Let $x_{\part{s}}$ be a block that does not change during the estimation from time $t_0$ to $t_1$. For this block
\begin{align}
 {\mu_x}_{\part{s},t_0}&={\mu_x}_{\part{s},t_1}={\mu_{\hat{x}}}_{\part{s},t_0}={\mu_{\hat{x}}}_{\part{s},t_1}\\
 P_{{x}_{\part{s}}{x}_{\part{s},t_0}}&=P_{{\hat x}_{\part{s}}{\hat x}_{\part{s},t_0}} = P_{{x}_{\part{s}}{x}_{\part{s},t_1}}=P_{{\hat x}_{\part{s}}{\hat x}_{\part{s},t_1}} =P_{{\hat x}_{\part{s}}{x}_{\part{s},t_1}}
\end{align}
and there is no need to compute their values before the final update where the whole state is updated. We can see by looking at \eqref{equ:muv} and \eqref{equ:Rv} that the formulation presented in \cite{guivant_2017} would require inversion of zero matrices.

In \eqref{equ:covup} the inverses can be computed blockwise. Because  
block $\part{s}$ of $\PhI$ and $\PdI$ are identical their inverses cancel each other and so do not need to be computed. In the virtual state propagation $F_{\part{s}}^v$ is an identity matrix and $Q_{\part{s}}^v$ is a zero matrix. 
An example of applying the block algorithm is presented in Section~\ref{sec:iteKFE}.

\section{Example Applications}
\label{sec:examples}

\subsection{\uus{sec1}{Fourier series}}
\label{sec:fourier}
In this example, we show how the general form of Algorithm~\ref{algo:partiallinearup} is simplifed for a specific application and compare the computational complexity of the optimized version when we are using \ac{UKF} as the estimation algorithm. In this example an $\statedim$ dimensional state that is inferred given a scalar measurement of form
\begin{equation}
	y = a_0+\sum_{j=1}^k\left( a_j \sin ( j x_1) + b_j \cos ( j x_1)\right) + \measnoise,
\end{equation}
where $\measnoise \sim \N (0,R)$ is independent of the state. The augmented state is $\smatr{x \\ \measnoise}$. Matrices for Algorithm~\ref{algo:partiallinearup} are $T=\smatr{1 & 0 & \ldots}$ and $H=\smatr{0&\ldots& 0& 1}$. Using these the transformed mean is $\mean_{\tilde z} = \mean_{x_1}$ and the transformed covariance is $P_{\tilde z\tilde z} = P_{x_1x_1}$. Sigma-points are  generated using one-dimensional mean $\mean_{x_1}$ and covariance $P_{x_1x_1}$ and the moments are
\begin{equation}
\begin{aligned}
	\mu_{g( \tilde z)} & = \sum  w_i g(\sigmap_i) \\ & = \sum \left( w_i a_0+\sum_{j=1}^k\left( a_j \sin ( j \sigmap_i) + b_j \cos ( j \sigmap_i)\right) \right) \\
	P_{g( \tilde z)g( \tilde z)} &= \sum w_i (g(\sigmap_i) -  \mu_{g( \tilde z)} )(g(\sigmap_i) -  \mu_{g( \tilde z)} )^T \\ 
	P_{\tilde z g( \tilde z)} &= \sum w_i (\sigmap_j - \mean_{\tilde z})(g(\sigmap_i) -  \mu_{g( \tilde z)} )^T.  
\end{aligned}
\end{equation}
and equation \eqref{equ:replace} is
\begin{equation}
P_{z g(\tilde z)} =  P_{zz[:,1]}P_{zz[1,1]}^{-1}P_{\tilde z g( \tilde z)} \label{equ:dsr}
\end{equation}
Because measurement error has zero mean we have
\begin{equation}
 \mu_{h(z)} = \mu_{g( \tilde z)}
 \end{equation}
and, as it is  independent of the state, the last element of \eqref{equ:dsr} is 0. Thus, we can simplify  \eqref{equ:isoP} to 
\begin{equation}
P_{h(z)h(z)} = P_{g(\tilde z)g(\tilde z)} + R
\end{equation}
and \eqref{equ:pxh} to
\begin{equation}
 P_{xh(z)} = P_{zz[:,1]}P_{zz[1,1]}^{-1}P_{\tilde z g( \tilde z)}.
 \end{equation}

Using the above equations we evaluate the complexity improvements when they are applied with \ac{UKF} that uses $2n+1$ sigma-points.  Table~\ref{tbl:complexity} shows the asymptotic complexities of application of basic~\ac{UKF} and optimized version. The complexities are given as functions of $k$ to take into account the complexity of the measurement function and of $n$, the number of state variables. The highest complexities are underlined.

\begin{table}
\caption{Computational complexity of parts of the update of the state with different optimizations}
\label{tbl:complexity}
\centering
\begin{tabular}{r|c|c} 
Computation of & Optimized version & Basic \ac{UKF}\\ \hline
	Sigma points & $\mathcal{O}(1)$ & \underline{$\mathcal{O}(n^3)$}\\
	$\mu_{g(z)}$ & $\mathcal{O}(k)$ & $ \mathcal{O}(nk) $  \\ 
	$P_{h(z)h(z)}$ & $  \mathcal{O}(k)$ & $  \mathcal{O}(nk)$ \\
	$P_{zg(z)} $ &  $ \underline{ \mathcal{O}(n+k) }$ & \underline{$\mathcal{O}(n^2+nk)$} \\ 
	$K$ & $\mathcal{O}(n)$ &  $\mathcal{O}(n)$ \\
	$\mu$ & $\mathcal{O}(n)$ & $\mathcal{O}(n)$ \\
	$P$ & \underline{$\mathcal{O}(n^2)$} & $\mathcal{O}(n^2) $  \\
\end{tabular}
\end{table}

For the optimized version, the most complex parts of the update have complexities $ \underline{ \mathcal{O}(n+k) }$ for computing the cross correlation and $ \underline{\mathcal{O}(n^2)}$ for updating the covariance matrix. The most demanding parts for basic \ac{UKF} are also computing the cross correlation which has complexity \underline{$\mathcal{O}(n^2+nk)$} and the computation of sigma points which has complexity \underline{$\mathcal{O}(n^3)$} due to computing the matrix square root of the covariance matrix. This implies that the given optimization reduces computational load in both situations where $n$ or $k$ dominates. 

\subsection{Pedestrian Dead Reckoning}
\label{sec:pdr}
In \ac{PDR}, a pedestrian's position is estimated based on steps and heading changes that are measured with accelerometers and gyroscopes. In \cite{nurminen2013a}, a state consists of two-dimensional position $r$, heading~$\theta$, step length $l$, and possibly a floor index. The state transition is computed with a particle filter. Here we show how the state model without floor index and with normal  noise can be optimized for use with \acp{KFE}. We use in this example algorithms presented in Sections~\ref{sec:partlin} and~\ref{sec:condlin}.

The nonlinear state transition model is
\begin{equation}
\begin{aligned}
	x_{t+1} & = \smatr{r_{1,t+1} \\ r_{2,t+1} \\\theta_{t+1} \\ l_{t+1}  } \\
	       & =
	\smatr{r_{1,t}  + \left(l_{t}+\noise_{l}^x \right) \cos \left(\theta_t + \noise_\theta^x \right)+ \noise_{r,1}^x \\
	          r_{2,t}  + \left(l_{t}+\noise_{l}^x \right) \sin \left(\theta_t + \noise_\theta^x \right) + \noise_{r,2}^x \\
	          \theta_{t} + \noise^x_\theta \\
	          l_t + \noise^x_l
	          }, \label{equ:fullpdr}
	          \end{aligned}
\end{equation}
where  $\noise^x_\theta \sim \N(\Delta_t, \sigma^2_\theta)$ is the noisy heading change, with $\Delta_t$ measured from gyroscopes, $l$ is the footstep length and $r$ is the position of the pedestrian. The augmented state is $z= \smatr{r_1 & r_2 & \theta & l & \noise^x_{r,1} & \noise^x_{r,2} & \noise^x_\theta & \noise^x_l}^T$. The state transition model can be written  in the form of \eqref{equ:fullequ}:
\begin{equation}
\begin{aligned}
	\tilde z & = \smatr{\theta_t + \noise^x_\theta \\ l_{t}+\noise^x_{l}   } \\
	T &= \smatr{\mathbf{0}_{2 \times 2} & I_{2 \times 2} & \mathbf{0}_{2 \times 2} & I_{2 \times 2}  } \\
	         	g(\tilde{z}) & = \smatr{ \tilde{z}_2 \cos \tilde{z}_1 \\  \tilde{z}_2 \sin \tilde{z}_1 } \label{equ:reducedpdr} \\
		A &= \smatr{I_{2 \times 2} \\ \mathbf{0}_{2 \times 2} } \\
	H & = \smatr{I_{4 \times 4}  I_{4 \times 4} }.
\end{aligned}
\end{equation}
In this formulation the reduced state $\tilde z$ is two-dimensional and the nonlinear part of the state transition function is also two-dimensional.

If this problem is to be solved with a \ac{KFE} that uses sigma-points, we can optimize the implementation further. This can be done because $\tilde{z}_2$ in \eqref{equ:reducedpdr} is conditionally linear, given $\tilde{z}_1$. The parts of function \eqref{equ:condl}  are
\begin{equation}
\begin{aligned}
	z_n &= \tilde{z}_1 \\  z_l &= \tilde{z}_2\\ 
	g_n(\tilde{z}_1) &= \smatr{ 0\\0 } \\ 
	G_n(\tilde{z}_1) &=\smatr{ \sin \tilde{z}_1\\\cos \tilde{z}_1}.
\end{aligned}
\end{equation}
Using this the dimension of the nonlinear part of the state is reduced to 1 and the sigma-points are generated for only one dimension. 

The reduction of the computational complexity using these optimizations depend on the used \ac{KFE}. Using \ac{QKF} with parameter $\alpha=3$ the number of sigma-points for computing \eqref{equ:UKF1}-\eqref{equ:UKF3} is reduced from 6561 to 3. With  \ac{UKF} the reduction is not as significant,  the number of sigma-points is reduced from 17 to 3. 

\begin{figure}
\centering
\includegraphics[width=\columnwidth]{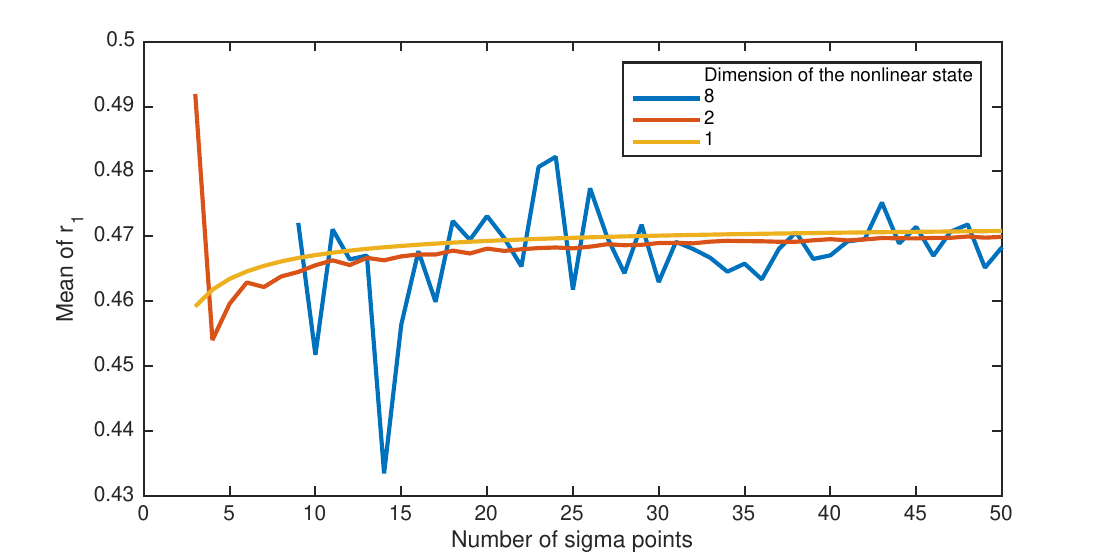}
	\caption{Estimated mean of $r_1$ with different number of function evaluation points using \ac{S2KF}}
	\label{fig:stepex}
\end{figure}

Figure~\ref{fig:stepex} shows mean estimates of the first position variable $r_1$ after one propagation step in a simulation. The initial state has step length 1 and step direction $50\frac{\pi}{180}$ degrees. The initial covariance is strongly correlated being $\diag\left( 1,1,1,0.01 , .01 ,.01, \frac{\pi^2}{180^2}, .00001 \right) +\mathbf{1} $ . 
The estimates are computed using \ac{S2KF} with a varying number of sigma-points. An implementation of the sigma-point generator and \ac{S2KF} can be found in \cite{nonlinearestimationtoolbox}.  \ac{S2KF} is applied to the original state \eqref{equ:fullpdr} and to the above presented reduced states with 2 and 1 nonlinear variables. The propagation of the state with the full state model cannot be computed with fewer than 9 sigma-points.

The state computed using the model that uses conditional linearity in addition to the linearity has the smoothest convergence to an estimate and the estimate with full state estimate has the largest variation.  This shows that the use of optimizations can also improve the estimation accuracy as the sigma-points will be better aligned along the dimensions with nonlinearities. However, if one would use a \ac{KFE} whose number of sigma-points cannot be altered, such as having $2\statedim +1$ sigma-points in \ac{UKF}, one may end up with better (or sometimes worse) accuracy without optimizations just because there are more sigma-points.  We don't compute the accuracy of the estimates in the following examples as it is dependent on the selected \ac{KFE} and parameters. One can expect the accuracy to be approximately the same when the optimizations are applied.

\subsection{\uus{sec2}{Remaining useful life prediction of lithium-ion battery}}
\label{sec:lithium}
In this example we apply the optimizations to prediction of the useful lifetime of a lithium-ion battery. The original model is proposed in \cite{MIAO2013805}.

The state consists of four estimated variables $x=\matr { x_1 & x_2 & x_3 &x_4}^T$ and the state transition model is linear $x_{t+1} = x_{t} + \statenoise$, where $\statenoise \sim \N(0,Q)$  and measurement model is 
\begin{equation}
	y_t = x_{3,t} e^{t x_{1,t} }+x_{4,t} e^{t x_{2,t} } + \measnoise,
\end{equation}
where $y_t$ is the measured capacity of the lithium-ion battery at cycle $t$, and $\measnoise$ has variance $R$.

Because the state transition model is completely linear it is evident that the propagation should be done using the linear \ac{KF}, which becomes
\begin{equation}
\begin{aligned}
	\mean_{x_t}^- & = I \mean_{x_{t-1}} = \mean_{x_t} \\
	\statecov_{x_tx_t}^- & = I \statecov_{x_{t-1}x_{t-1}} I^T + Q_t  = \statecov_{x_{t-1}x_{t-1}} + Q_t
\end{aligned}
\end{equation}
This prediction requires 16 additions when $Q$ is added to the state covariance, or $4$ if $Q$ is diagonal. If the prediction were computed with \ac{UKF} and an augmented state, there would be 11 sigma-points with 5 elements each, thus computation of predicted covariance \eqref{equ:UKF2} requires 550 products and 275 summations. In addition, generation of sigma points involving matrix square root computation and computation of the mean \eqref{equ:UKF1} have to be performed. The absolute numbers of computations are small for modern computers either way, but reducing the number of computations by a factor greater than 200 is a significant reduction of operations. 

The measurement model has one additive component, $\measnoise$, and we can use \eqref{equ:U1}-\eqref{equ:U3} for it. Variables $x_3$ and $x_4$ are conditionally linear and we can apply equations from Section~\ref{sec:condlin}. In this case 
\begin{equation}
\begin{aligned}
	\mu_{x_l} &= \mu_{[3,4]}  \\
	\mu_{x_n} &= \mu_{[1,2]} \\ 
	P_{x_l x_l} & = P_{[3,4],[3,4]} \\	
	P_{x_l x_n} & = P_{[3,4],[1,2]} \\
	P_{x_n x_n} & = P_{[1,2],[1,2]} \\
	g_n(\sigmap_i) & = 0\\
	G_n(\sigmap_i) & = \matr{ e^{t \mu_1}  & e^{t \mu_2}} \\
\end{aligned}
\end{equation}
These can be used to compute $\mu_{x_l | \sigmap_i }$ \eqref{equ:condmu} and $P_{x_lx_l | \sigmap_i }$ \eqref{equ:condP}.
Merging \eqref{equ:U1}-\eqref{equ:U3} and \eqref{equ:condl1}-\eqref{equ:condl3} gives us
\begin{equation}
\begin{aligned}
	{\mathcal Y}_i  =& G_n(\sigmap_i)\mu_{x_l | \sigmap_i } \\
	\mu_{g(z)} = & \sum w_i  {\mathcal Y}_i\\
	P_{g(z)g(z)} = & \sum w_i \left(({\mathcal Y}_i -\mu_{g(z)})( {\mathcal Y}_i -\mu_{g(z)})^T  \right. \\ & + \left. G_n(\sigmap_i) P_{x_lx_l}G_n(\sigmap_i)^T \right)  + R \\
		P_{zg(z)} = & \sum w_i \left( \left( \matr{\sigmap_i \\ \mu_{z_l | \sigmap_i }}  - \mu_x \right)( {\mathcal Y}_i -\mu_{g(z)})^T \right.  \\  & + \left.    \smatr{0 \\P_{x_l x_l | \sigmap_i }\Mfun_n(\sigmap_i)^T  }  \right).
\end{aligned}
\end{equation}
In this situation, the computation gains come mostly from the reduction of the number of sigma-points as the dimension is reduced from 5 to 2. When using \ac{UKF} the number of sigma-points is reduced from 11 to 5. However, the above computations are more complex for each summand than if the moments would have been computed using \eqref{equ:UKF1}-\eqref{equ:UKF3}. So the optimizations based on conditional linearity are not reducing the computational complexity. However, when using \ac{QKF} with parameter $\alpha=4$, the number of sigma-points is reduced from $4^5=1024$ to $4^2=16$ and the optimizations are useful.

\subsection{Source Tracking Using a Microphone Array \ac{TDoA}}
\label{sec:tdoa}
In tracking of a sound source using a microphone array, the received acoustic signals of two microphones are compared so that the \ac{TDoA} of the acoustic signal is measured. In the ideal situation the \ac{TDoA} is
\begin{align}
\measfun_{i,j}(r_0) & = \frac{\norm{r_i - r_0 } -  \norm{r_j - r_0 }}{v},
\end{align}
where $r_0$ is the location of the sound source, $r_i$ is the location of the $i$th microphone and $v$ is the speed of sound. When there are $m$ microphones in the array there are at most $ \frac{1}{2}m(m-1)$ \ac{TDoA} measurements available  \cite{oualil2012tdoa}.

In practice,  measurements contain noise. Here we consider model with noises
\begin{align}
\measfun_{i,j}(x) & = {\norm{r_i - x } -  \norm{r_j - x } + \varepsilon^y_i - \varepsilon^y_j + \varepsilon^y_{i,j}},
\end{align}
where $\varepsilon_i$ is the noise corresponding to $i$th microphone and $\varepsilon_{i,j}$ is the error corresponding to the correlation of the measured signals from the microphone pair. The speed of sound is multiplied out from the equation.

\uus{noisecov}{When $\varepsilon_{i,j}=0$ for all microphone pairs the measurement noise covariance is not full rank and the measurement equations can be modeled with $m-1$ measurements of the form
\begin{align}
\measfun_i(x) & =   \norm{r_i - x } -   \norm{r_m - x } + \varepsilon^y_i - \varepsilon^y_m,
\end{align}
which has a full rank noise covariance matrix.} This kind of assumption is done for example in \cite{bechler2004system}. In practice, this assumption does not hold and the selection of microphone pairs is a tradeoff, where pairs close to each other have smaller correlation error $\varepsilon_{i,j}$, but worse geometry than pairs far away from each other~\cite{silverman1998huge}.
Here we consider the situation where all possible microphone pairs are used and errors $\varepsilon_i$ and $\varepsilon_{i,j}$ are modeled  as Gaussians. The augmented state model is 
\begin{equation}
z=\smatr{x^T & \varepsilon_{1,1}^y & \varepsilon_{1,2}^y & \ldots & \varepsilon_{m,m}^y & \varepsilon_{1}^y & \ldots &\varepsilon_{m}^y }^T,
\end{equation}
where $x$ contains 3 position and velocity variables.

Using \eqref{equ:U1}--\eqref{equ:U3} the nonlinear part of the estimation can be done only for the state variables and  sigma-points would be required only for the 6 dimensional state instead of using sigma-points also for $\frac{1}{2}m(m+1)$ noise terms. The measurement model can be written in compact form using \eqref{equ:fullequ}:
\begin{equation}
\begin{aligned}
\tilde z & = x_{1:3} \\
T & = \smatr{I_{3\times3} & \mathbf{0_{3\times3}}} \\
g(\tilde z) & = \smatr {\norm{r_1 - \tilde z } &  \ldots& \norm{r_m - \tilde z }}^T \\
A & = \smatr{ \mathbf{1}_{m-1\times 1} & \hspace{1cm}-I_{m-1\times m-1} \hspace{-1cm}\\ 
 	       \mathbf{0}_{m-2\times 1}  & \mathbf{1}_{m-2\times 1 }  & -I_{m-2 \times m-2} \\
	       \mathbf{0}_{m-3\times 2}  & \mathbf{1}_{m-3\times 1 }  & -I_{m-3 \times m-3} \\
	        &\vdots & \\
	       \mathbf{0}_{1\times m-2}  & 1  & -1  } \\
H&=	 \smatr{ \mathbf{0}_{  \frac{m(m-1)}{2}\times 6  } & A & I_{ \frac{m(m-1)}{2} \times  \frac{m(m-1)}{2}}}.
\end{aligned}
\end{equation}
Using these the dimension of the nonlinear function is reduced from $\frac{1}{2}m(m-1)$ to $m$. This reduces the number of updated elements for each sigma-point in \eqref{equ:UKF2} from $\left(\frac{m(m-1)}{2}\right)^2$ to $m^2$. $A$ and $H$ matrices are sparse and an application of sparse linear algebra codes would further enhance the performance of the algorithm.
 
The dimension of the innovation covariance $S$ is $\left(\frac{m(m-1)}{2}\right)^2 \times \left(\frac{m(m-1)}{2}\right)^2$. The computation of the inverse of the innovation covariance for Kalman gain \eqref{equ:Kalmangain} can be optimized using the inversion formula \eqref{equ:woodbury}. The noise terms are assumed independent so the noise covariance matrix $R$ is diagonal. We partition noises into two parts, and denote the covariance matrices  $D_1$ and $D_2$. $D_1$ corresponds to microphone specific terms $\measnoise_i$ and $D_2$ to microphone pair specific terms $\measnoise_{i,j}$. Because $\tilde z$ contains only state variables and they do not contribute to the linear part ($AP_{zg(\tilde z)^T}H^T=\mathbf{0}$), the innovation covariance can be written as 
\begin{equation}
	S = D_2 + A( P_{g(\tilde z)g(\tilde z)} + D_1) A^T 
\end{equation}
and its inverse is
\begin{equation}
\begin{aligned}
	S^{-1} = &  D_2^{-1} + D_2^{-1}A Q^{-1} A^T D_2^{-1},
	\end{aligned}
\end{equation}
where $Q=\left[( P_{g(\tilde z)g(\tilde z)} + D_1 )^{-1} + A  D_2^{-1}A^T  \right]$. Because $D_2^{-1}$ is diagonal its inverse is diagonal with the reciprocal of the diagonal elements of $D_2$ on its diagonal. Other inverses that appear in this formula are applied to $m\times m$ matrices instead of $\left(\frac{m(m-1)}{2}\right)^2 \times \left(\frac{m(m-1)}{2}\right)^2$ matrices. If the  used matrix inversion algorithm has complexity of $\mathcal{O}{(k^3)}$, the complexity of the inversion operation is reduced from $\mathcal{O}{(m^{12})}$ to $\mathcal{O}{(m^3)}$.

\subsection{Optimization of Iterative \acp{KFE} }
\label{sec:iteKFE}
The optimizations based on the division of the state into multiple sub-blocks in Section~\ref{sec:multiprocess} are most widely used in the field of \ac{SLAM} \cite{938382,guivant_2017,estrada2005hierarchical}. In \ac{SLAM}, the state contains a dynamic part that represents the agent (robot) that is mapping static landmarks in the environment and localizing itself. The landmark coordinates are the static part of the state. 

We propose that these optimizations can be applied also to \acp{KFE} that do the update iteratively e.g.\ \ac{RUF} \cite{RUKF} and its extension for sigma-point filters \cite{RUFEXT} or posterior linearization filters \cite{IPLF,8291501}. In~\ac{RUF}, the state is updated by applying the update in parts that have smaller impact than the original update. After each part of the update the linearization is recomputed. These iterative filters are developed assuming additive Gaussian noise. Instead of making the updates for the full state the update could be done iteratively only for the observed variables. Because the iterative update is actually for a single observation, the unobserved state variables are static during the partial updates.

\begin{figure}[tb!]
\centering
\includegraphics[width=0.8\columnwidth,clip=true, trim=0cm 1.1cm 0cm 0.7cm]{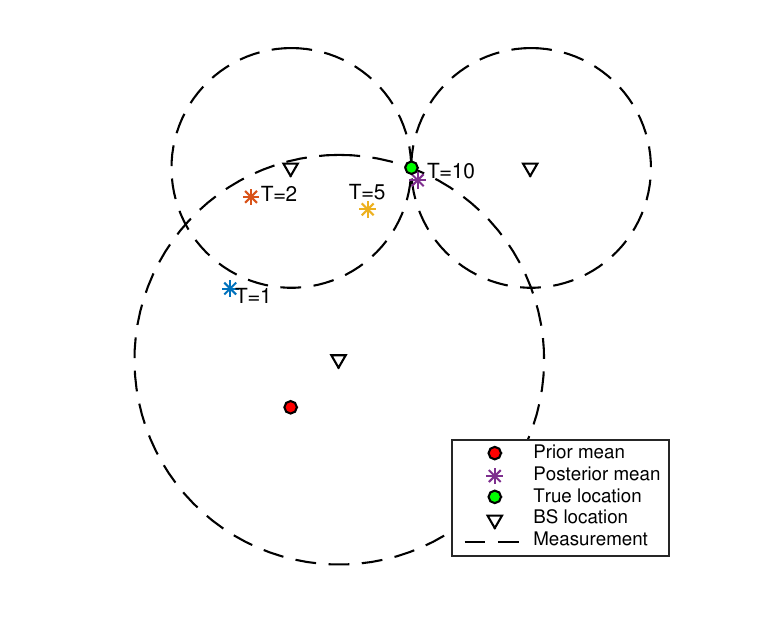}
\caption{Posterior means computed with \ac{RUF} with varying number of iterations.}
\label{fig:rangex}
\end{figure}

In this example we consider a state model with 3D position, velocity, and acceleration i.e.\ 9 state variables. Measurements used are range measurements from 3 base stations. The measurement model for the range from the $i$th base station is 
\begin{equation}
	y_i =  \left\| x_{[1:3]}-  r_i \right\| + \measnoise_i, \label{equ:rangemeas}
\end{equation}
where $r_i$ is the location of the $i$th base station. In our example, the prior has a large variance and its mean is chosen so that the linearization about the mean is not accurate. Figure~\ref{fig:rangex} shows the example situation and the posterior means which are computed with different number of iterations of \ac{RUF}. The estimate with one iteration ($T=1$) is identical to the \ac{EKF} estimate and is not close to the true location. The estimate with 10 iterations is close to the true location. The computation of 10 iterations with \ac{RUF} involves the updates to the $9\times 9$ state covariance matrices 10 times. Because the measurement model \eqref{equ:rangemeas} depends only on the three first state variables this block can be updated 10 times using methods presented in Section~\ref{sec:multiprocess} and the remaining variables can be left in a block that does not need any update until the full covariance update. The full covariance update is computed after all iterations are applied. In this update scheme, the $3\times 3$ covariance matrix is updated $T$ times and the $9\times9$ matrix is updated only once using {\eqref{equ:unobstart}--\eqref{equ:unobend}}. Thus, in each iteration the covariance update is 9 times faster. Moreover, if the \ac{KFE} computes sigma-points using Cholesky decomposition, as \ac{UKF} does, there is more gain in speed. This is because the complexity of Cholesky decomposition is $\mathcal{O}(\statedim^3)$.

\section{Conclusions}
\label{sec:conclusions}

In this tutorial, we presented different optimizations for \acp{KFE} that exploit the structure of the state and measurement models to reduce the computational load. Table~\ref{tbl:optitaulu} summarizes the structures of models that are exploited in different optimizations. These structures are present in various nonlinear estimation problems.

\begin{table}[tb]
\caption{Summary of presented optimizations} \label{tbl:optitaulu} 
\footnotesize
\centering
{\begin{tabular}{c|p{6.2cm}}
Section & Exploited structure \\ \hline
Section~\ref{sec:partlin} & Partially linear functions ${f(z) = A g(Tz) + Hz}$ \\
Section~\ref{sec:condlin} & Conditionally linear functions ${f(z) = g_n(z_n) + G(z_n)z_l}$ \\
Section~\ref{sec:Innopti} & Measurement covariance is block diagonal \\
Section~\ref{sec:innopti2} &Innovation covariance of form ${S = D_2 + A( P_{g(\tilde z)g(\tilde z)} + D_1) A^T}$ \\
Section~\ref{sec:multiprocess} & State can be divided into multiple subprocesses \\
\end{tabular}
}	
\end{table}
Table~\ref{tbl:Kalmantaulu} shows some properties of selected \acp{KFE}.  Different optimizations give different amounts of speed increase with different \acp{KFE}. Optimizations presented in sections \ref{sec:partlin} and \ref{sec:condlin} are most useful with \acp{KFE} that have a high number of function evaluations as a function of state dimension. The exploitation of conditionally linear part of functions in Section~\ref{sec:condlin} requires that the expectations are approximated with equations of form \eqref{equ:UKF1}--\eqref{equ:UKF3}. 

\begin{table}[tb]
\caption{Summary of properties of selected \acp{KFE}} \label{tbl:Kalmantaulu}
\footnotesize
\centering
	{\begin{tabular}{c|cccc}
	 Algorithm & \rotatebox{90}{\parbox{4cm}{Order of measurement function evaluations \\ $\alpha$ = algorithm specific parameter ($\alpha>0$)\\ $\iterations$ = number of iterations}}  & \rotatebox{90}{\parbox{4.4cm}{Computation of $\jacobian$ \\ A=analyticall \\ I=using \eqref{equ:virtJac}\\ N=numerically inside algorithm }} & \rotatebox{90}{Expectations of form \eqref{equ:UKF1}--\eqref{equ:UKF3}} &\rotatebox{90}{Iterative algorithm}  \\ \hline
	 \ac{EKF} \cite{jazwinski}& 1 & A & \\
	 \ac{EKF2} \cite{jazwinski} & 1 & A & \\
	 \ac{UKF} \cite{UKF}& \statedim  &  I & X &\\
	 \ac{CKF} \cite{cubature}& $\statedim^\alpha$ & I & X &  \\	 
	 \ac{S2KF} \cite{S2KF}& $\statedim+\alpha$ & I & X &  \\	 
	 \ac{GF} \cite{huber2008gaussian}& $\alpha\statedim$ & I & X &  \\	 
	 \ac{QKF} \cite{ITO}& $\alpha^\statedim$ & I & X &  \\	 
	 \ac{DD} \cite{norgaard}& $\statedim$ &N & \\
	 \ac{SOKF2} \cite{ITO}& $\statedim^2$ & N& \\
	  \ac{RUF} \cite{RUKF} & $\iterations$ & A & &X \\
%	  \ac{URUF} \cite{RUFEXT} & $\iterations n$ & I & X &X \\
	\end{tabular}}
\end{table}

In addition to surveying existing code optimization methods and using a general unified notation that allows them to be used with various \acp{KFE}, we make the following contributions: 
\begin{enumerate}
	\item We point out the possibility to use a linear transformation to find the minimal nonlinear subspace (Section~\ref{sec:partlin}).
	\item We introduce an algorithm for systems with conditionally linear states (Section~\ref{sec:condlin}); it can be used 
to compute the moments for the nonlinear part and solve others using linearity (Section~\ref{sec:partlin}) as is done in the example in Section~\ref{sec:pdr}.
	\item We present a new formulation for estimation with a state that can be divided into separate estimation processes (Section~\ref{sec:multiprocess}). The new formulation avoids dealing with inverses of singular matrices.
	\item We present an example showing that in addition to gaining increase in speed of computation the optimization algorithms may also lead to better estimation accuracy (Section~\ref{sec:pdr}).
	\item We show how to use the matrix inversion lemma to reduce the computational complexity in \ac{TDoA}-based positioning (Section~\ref{sec:tdoa}).
	\item  We present an example showing that the optimization that exploits a static state and partly unobserved variables can be applied with iterative \acp{KFE} (Section~\ref{sec:iteKFE}).
\end{enumerate}

Optimizations in this tutorial were applied to \acp{KFE} that use a mean vector and covariance matrix to represent the state. However, there is another flavour of \acp{KF}, called square root filters, that propagate a square root of the covariance matrix instead of the full matrix for better numerical properties  \cite{thornton,1977fmd,SQKF,940586}. In \cite{5670900}, square root filtering is optimized for the situation where some of the state variables are linear and applying independent measurements sequentially as in Section~\ref{sec:Innopti} is  straightforward. However, most of the optimizations presented in this tutorial have not been considered for square-root form in literature and this remains an open topic.

Another interesting topic is whether optimizations such as the ones in this tutorial could be applied with \ac{EnKF}. \ac{EnKF} is mainly used for problems whose state space dimension is so large that it is infeasible to work with the covariance matrix. In such a context, one can expect that the effort invested in finding and implementing computation optimizations would be especially worthwhile.

\balance
%\section*{References}
\bibliographystyle{acm}
\bibliography{viitteet_rbkfe2}

%\clearpage
%\begin{acronym}
%\acro{CKF}{Cubature Kalman Filter}
%\acro{DD}{Divided Difference}
%\acro{EKF}{Extended Kalman Filter}
%\acro{EKF2}{Second Order Extended Kalman Filter}
%\acro{GF}{Gaussian Filter}
%\acro{GMF}{Gaussian Mixture Filter}
%\acro{KF}{Kalman Filter}
%\acro{KFE}{Kalman Filter Extension}
%\acro{PDR}{Pedestrian Dead Reckoning}
%\acro{QKF}{Gauss-Hermite Quadrature Kalman filter}
%\acro{RBPF}{Rao-Blackwellized Particle Filter}
%\acro{RUF}{Recursive Update Filter}
%\acro{SLAM}{Simultaneous Localization and Mapping}
%\acro{SOKF2}{Second Order Polynomial Kalman Filter}
%\acro{S2KF}[S\textsuperscript{2}KF]{Smart Sampling Kalman Filter}
%\acro{TDoA}{Time Difference of Arrival}
%\acro{UKF}{Unscented Kalman Filter}
%\acro{URUF}{Unscented Recursive Update Filter}
%\end{acronym}
\clearpage
\onecolumn
\appendix

Derivation of the virtual update that does not require the inversion of a possibly singular matrix. First, we define matrices $P_{\hat{x}\hat{x},t_1}$ and $P^{D}_{\hat{x}\hat{x},t_0}$
\begin{equation}
	\Ph = \matr{ P_{\hat{x}_{\part{1} }\hat{x}_{\part{1}},t_1}  & \mathbf{0} & \ldots  & \\
	                                                \mathbf{0} &  P_{\hat{x}_{\part{2} }\hat{x}_{\part{2}},t_1}  & \mathbf{0} & \ddots \\
	                                                \vdots & \mathbf{0} & P_{\hat{x}_{\part{3} }\hat{x}_{\part{3}},t_1}  & \ddots \\
	                                                 & \ddots & \ddots & \ddots} 
\end{equation}
\begin{equation}
	\Pd = \matr{ P_{{x}_{\part{1} }{x}_{\part{1}},t_0}  & \mathbf{0} & \ldots  & \\
	                                                \mathbf{0} &  P_{{x}_{\part{2} }{x}_{\part{2}},t_0}  & \mathbf{0} & \ddots \\
	                                                \vdots & \mathbf{0} & P_{{x}_{\part{3} }{x}_{\part{3}},t_0}  & \ddots \\
	                                                 & \ddots & \ddots & \ddots} 
\end{equation}
The virtual update covariance $R_{t_1}^v$ can be written using these and the matrix inversion lemma
\begin{equation}
\begin{aligned}
	\Rv & =  \Pd\left( \Pd - \Ph\right)^{-1}\Pd - \Pd \\
	&= \left( \PhI-\PdI \right) ^{-1}
\end{aligned}
\end{equation}
The predicted measurement is
\begin{equation}
\begin{aligned}
	y^v_{t_1} &= \mu_{x,t_0} + \Pd \left( \Pd - \Ph \right)^{-1}\left( \hat{\mu}_{t_1} - \mu_{x,t_0} \right) \\
	& =  \mu_{x,t_0} + \Pd \left( \PdI - \PdI\left(\PdI - \PhI\right)^{-1}\PdI \right) \left( \hat{\mu}_{t_1} - \mu_{x,t_0} \right) \\
	&=  \mu_{x,t_0} + \left( I - \left(\PdI - \PhI\right)^{-1}\PdI \right)  \left( \hat{\mu}_{t_1} - \mu_{x,t_0} \right) \\
	&=\hat{\mu}_{t_1} - \left(\PdI - \PhI\right)^{-1}\PdI  \left( \hat{\mu}_{t_1} - \mu_{x,t_0} \right)  
\end{aligned}
\end{equation}

The Kalman gain is
\begin{equation}
\begin{aligned}
	K_{t_1}^v&  = \Pm\left( \Pm + \RvI  \right)^{-1} \\ 
		        &= \Pm \left( \PmI -  \PmI \left(\RvI +   \PmI  \right)^{-1} \PmI \right) \\
		        &= I - \left(  \RvI +   \PmI  \right)^{-1} \PmI \\
		        &= \left(  \RvI +   \PmI  \right)^{-1} \left(  \RvI +   \PmI  \right) - \left(  \RvI +   \PmI  \right)^{-1} \PmI  \\ 
		        &= \left(  \RvI +   \PmI  \right)^{-1}   \RvI  \\ 
		        &=  \left(  \PhI - \PdI +   \PmI  \right)^{-1}   \left( \PhI-\PdI \right) 
\end{aligned}
\end{equation}

The posterior mean is
\begin{equation}
\begin{aligned}
\mu_{x,t_1}^v &= \mu_{x,t_0} + K_{t_1}^v \left( y^v_{t_1} - \mu_{x,t_0}  \right) \\ 
		 &=\mu_{x,t_0}  +  \left(  \PhI - \PdI +   \PmI  \right)^{-1}   \left( \PhI-\PdI \right) \\ 
		  & \times \left( \hat{\mu}_{x,t_1} - \left(\PdI - \PhI\right)^{-1}\PdI  \left( \hat{\mu}_{x,t_1} - \mu_{x,t_0} \right)  \right) \\
		  & =\mu_{x,t_0} + \left(  \PhI - \PdI +   \PmI  \right)^{-1}\\
		  & \times \left(   \left( \PhI-\PdI \right) \hat{\mu}_{x,t_1}  + \PdI  \left( \hat{\mu}_{x,t_1} - \mu_{x,t_0} \right)  \right) \\ 
		  & =\mu_{x,t_0} + \left(  \PhI - \PdI +   \PmI  \right)^{-1} \left( \PhI  \hat{\mu}_{x,t_1}  -\PmI \mu_{x,t_0} \right)
\end{aligned}
\end{equation}

The posterior covariance is
\begin{equation}
\begin{aligned}
	P_{xx,t_1}^v &= \Pm - K^v_{t_1} \left( \Pm + \RvI  \right) {K^v_{t_1}}^T \\
 	&= \Pm - \Pm \left( \PhI-\PdI \right) \left(  \PhI - \PdI +   \PmI  \right)^{-1}    \\
	&= \Pm \left(  \PhI - \PdI +   \PmI  \right) \left(  \PhI - \PdI +   \PmI  \right)^{-1} \\
	& - \Pm \left( \PhI-\PdI \right) \left(  \PhI - \PdI +   \PmI  \right)^{-1} \\
	& =\left(  \PhI - \PdI +   \PmI  \right)^{-1}. \label{equ:98}
\end{aligned}
\end{equation}

The posterior mean and covariance contain the inverses $\PhI$, $\PdI$ , $\PmI$,  and 
$\left(  \PhI - \PdI +   \PmI  \right)^{-1}$. If measurements have non-degenerate noise, the posterior $\PhI$ is positive definite. Matrices  $\PdI$ and $\PmI$ depend only on the prior which is assumed positive definite. The posterior has smaller or equal covariance than the prior, thus $\PhI - \PdI$ is positive semidefinite and the sum of this and the positive definite matrix $\PmI$ is positive definite. Thus, assuming that the prior covariance is positive definite and that the posterior is positive definite, the matrix inverses in~\eqref{equ:98} are applied only to positive definite matrices.

\begin{table*}[!b]
\begin{tabular}[t]{m{6cm}m{10cm}}
\vspace{5cm} \\
\includegraphics[width=6cm,clip=true,trim=7cm 0cm 7cm 0cm]{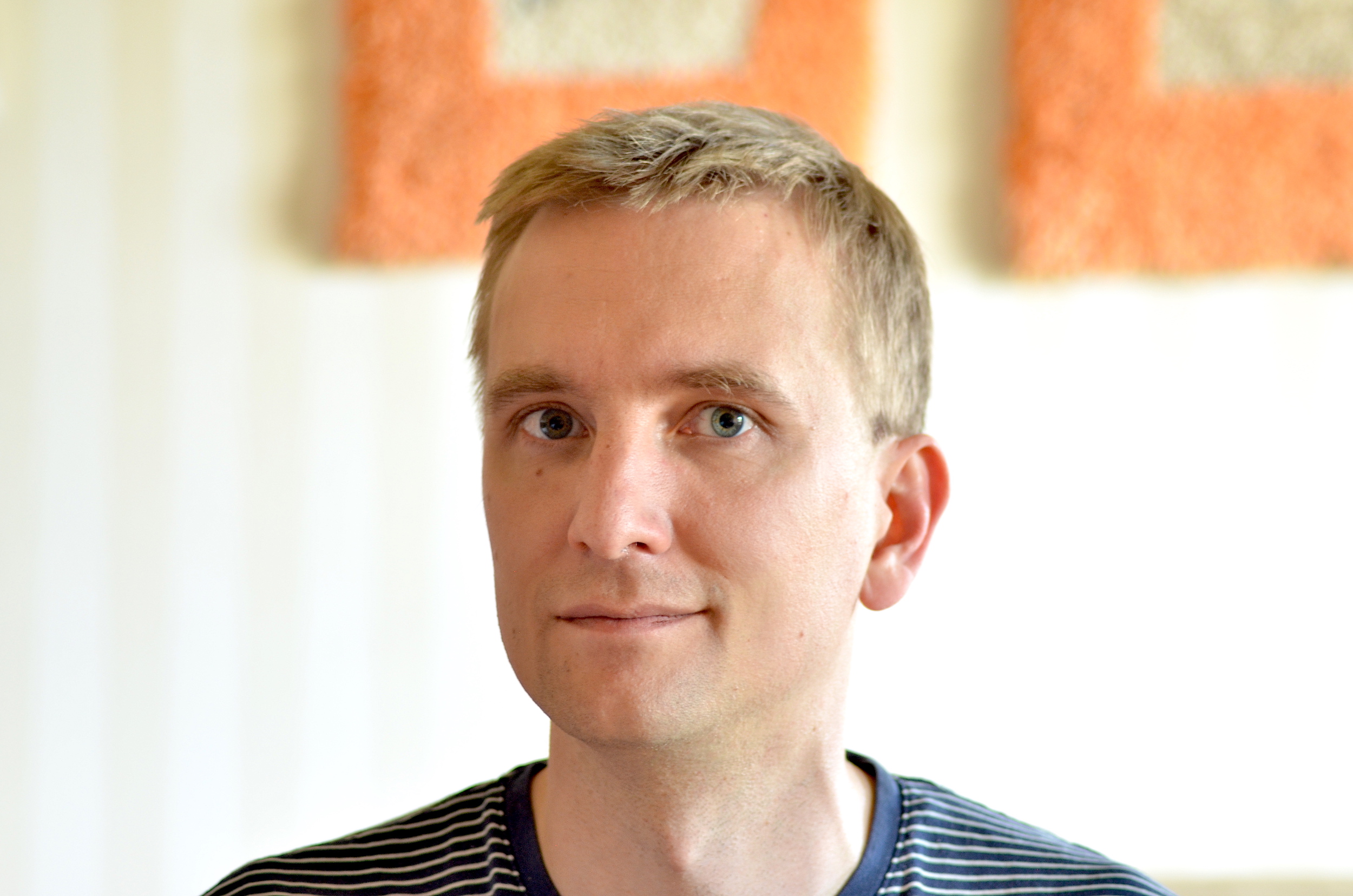}  & \normalsize Matti Raitoharju received M.Sc. and Ph.D. degrees in Mathematics in Tampere University of Technology, Finland, in 2009 and 2014 respectively. 
He works as a signal processing specialist at an aerospace and defence company Patria and as a postdoctoral researcher at Aalto University. He is a visiting scholar at Tampere University. His scientific interests include mathematical  modeling and development and application of Kalman type filters.\\ \vspace{1cm} \\
\includegraphics[width=6cm]{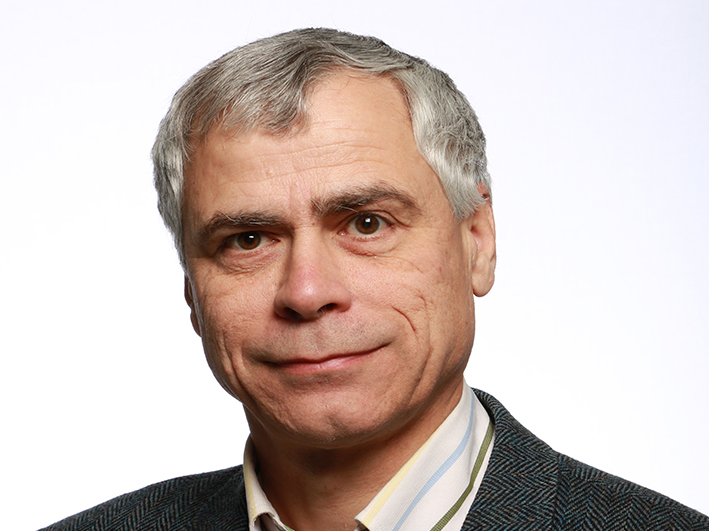} & \normalsize Robert Pich\'e received the Ph.D. degree in civil engineering in 1986 from the University of Waterloo, Canada. Since 2004 he holds a professorship in mathematics at Tampere University of Technology (now Tampere University). , Finland. His scientific interests include mathematical and statistical modelling, systems theory, and applications in positioning, computational finance, and mechanics.  
\end{tabular}
\end{table*}

\end{document}